\documentclass[
 reprint,
 amsmath,amssymb,
 aps,
prb
]{revtex4-2}

\usepackage{graphicx}
\usepackage{dcolumn}
\usepackage{bm}
\usepackage{gensymb}
\usepackage{amsmath}

\newcommand{\angstrom}{\mbox{\normalfont\AA}}

\begin{document}

\preprint{APS/123-QED}

\title{Theory of engineering flat bands in graphene using doubly-periodic electrostatic gating}%

\author{Nicholas Mario Hougland}
  \email{NMH61@pitt.edu}
  \affiliation{Department of Physics and Astronomy, University of Pittsburgh}
  \affiliation{Pittsburgh Quantum Institute, University of Pittsburgh}
\author{Ranjani Ramachandran}%
  \affiliation{Department of Physics and Astronomy, University of Pittsburgh}
  \affiliation{Pittsburgh Quantum Institute, University of Pittsburgh}
\author{Jeremy Levy}%
  \affiliation{Department of Physics and Astronomy, University of Pittsburgh}
  \affiliation{Pittsburgh Quantum Institute, University of Pittsburgh}
\author{David Pekker}%
  \affiliation{Department of Physics and Astronomy, University of Pittsburgh}
  \affiliation{Pittsburgh Quantum Institute, University of Pittsburgh}

\date{\today}

\begin{abstract}
%
We explore the use of applied electrical potentials to induce band flattening in graphene for bands near zero energy. We consider various families of doubly periodic potentials and simulate their effect on the electronic band structure using a tight-binding and a continuum approach. From these families, we find that an applied potential with symmetries of wallpaper group 17, in particular a Kagome potential, works best for inducing a high degree of band flattening for a range of realistic potential amplitudes and periods. Our work indicates that it should be possible to engineer the band structure of graphene using electrostatic gating, thus enabling a new approach to the development of graphene-based metamaterials.
\end{abstract}

\maketitle


\section{Introduction:}
Graphene has been found to exhibit numerous interesting and useful properties. It is a two-dimensional material and a semimetal with zero-gap due to Dirac points in its band structure, as first observed by Novoselov and co-workers~\cite{Wallace1947,Novoselov2004,Novoselov2005,Novoselov2005a,Geim2007}. In 2007, dos Santos and co-workers considered the effects of a small twist angle in bilayer graphene, using an effective model to find that the velocity near the Dirac points is reduced~\cite{LopesdosSantos2007}. This was further investigated when a tight-binding model also indicated the presence of flat bands~\cite{SurezMorell2010}. In 2011, Bistritzer and MacDonald found, using a continuum model, that at discrete magic angles, the Dirac velocity approaches zero~\cite{Bistritzer2011}. More recently, the experimental discovery of superconductivity and correlated insulator phases in twisted bilayer graphene (TBG) near the magic angle~\cite{Cao2018, Correlated2018, Lu2019,Sun2020,Yankowitz2019} has prompted interest in properties which can be induced in graphene, a field that has been named twistronics~\cite{Carr2017}. This magic angle, at approximately 1.1$\degree$, is where band flattening occurs near zero energy. It has since been shown that this band flattening is a fundamental feature of the magic angle~\cite{Tarnopolsky2019}, which raises the question of the role of flat bands in the phenomena seen with magic angle TBG (MATBG).

It has previously been predicted that applying a potential with periodicity in one direction to graphene would induce band flattening in one direction~\cite{Park2008}. More recently, this asymmetric band flattening was realized experimentally in graphene~\cite{Sudipta2013, Li2021}. In particular, it has been shown that band flattening can be induced, for instance, in the $k_x$ direction alone by applying a square wave potential along the $y$ spatial direction. Electrostatic gating of graphene with doubly-periodic potentials, i.e. 2D potentials with periodic structure in two different directions, like triangular and square lattices, has also been considered in the past. Indeed, it has been observed that lithographically defined gates can modify the electronic band structure~\cite{CheolHwan2008,Forsythe2018}.
%
%
Here, inspired by the possibility of using conductive AFM-lithography~\cite{huang2015, jnawali2018} or ultra-low-voltage electron-beam lithography~\cite{yang2020} to define arbitrary shaped electrostatic gates at complex-oxide interfaces, we ask whether doubly periodic potentials applied to monolayer graphene can induce symmetric band flattening as seen in twisted bilayer graphene.

\begin{figure}
\includegraphics[width=\columnwidth]{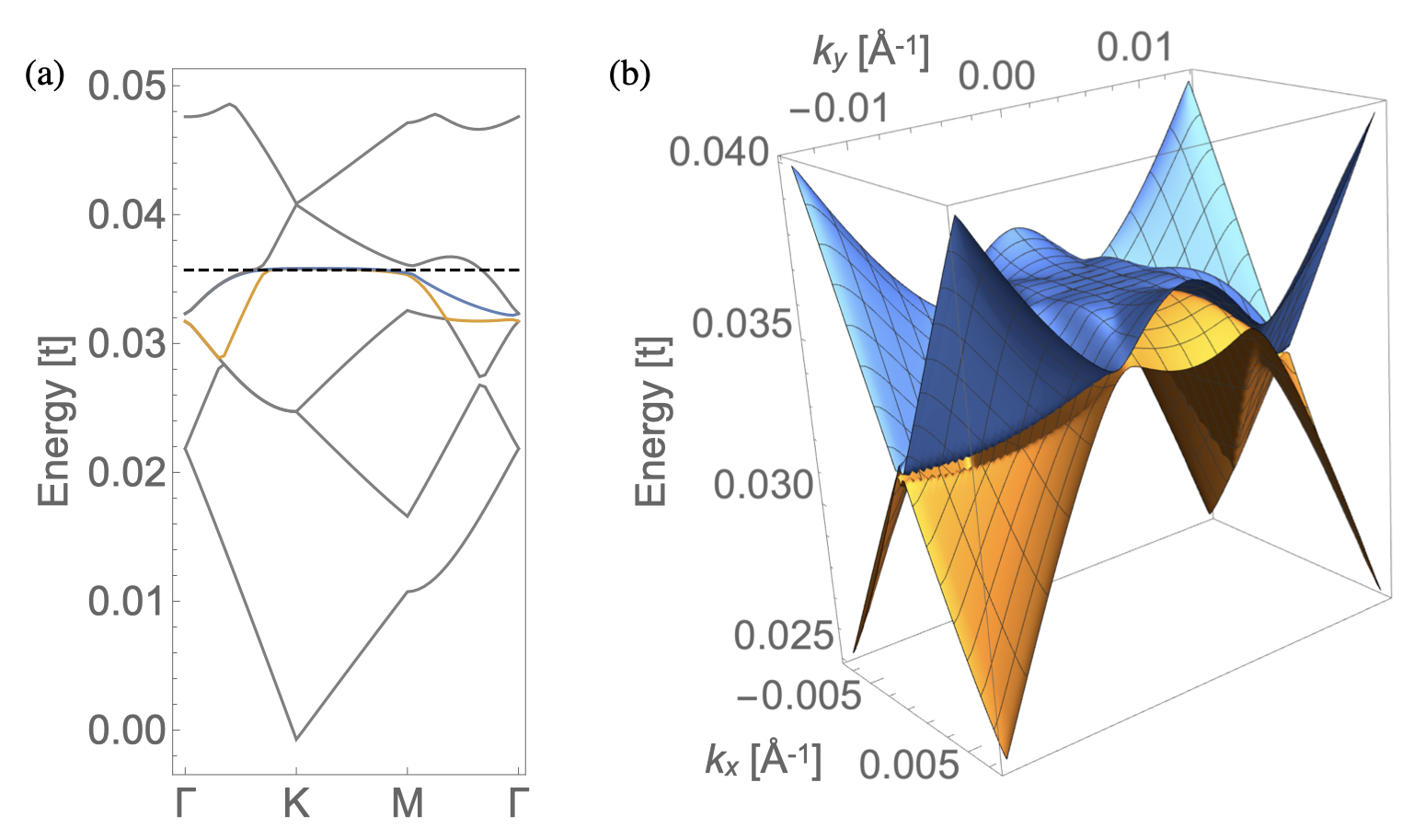}
\caption{\label{fig:plot3d} Isotropic band flattening induced by Kagome potential. The figure depicts two bands in the Brillouin zone of the superlattice (applied potential) obtained with the continuum model of gated monolayer graphene. At the point $k_x=k_y=0$, which corresponds to the location of the Dirac point in the ungated model, the Fermi velocity as well as the band curvature are both essentially zero. In part (a), the black line indicates a constant energy, demonstrating the flat nature of the highlighted bands. Part (b) shows the highlighted bands in both $k_x$ and $k_y$.}
\end{figure}

\begin{figure*}
\includegraphics[width=\textwidth]{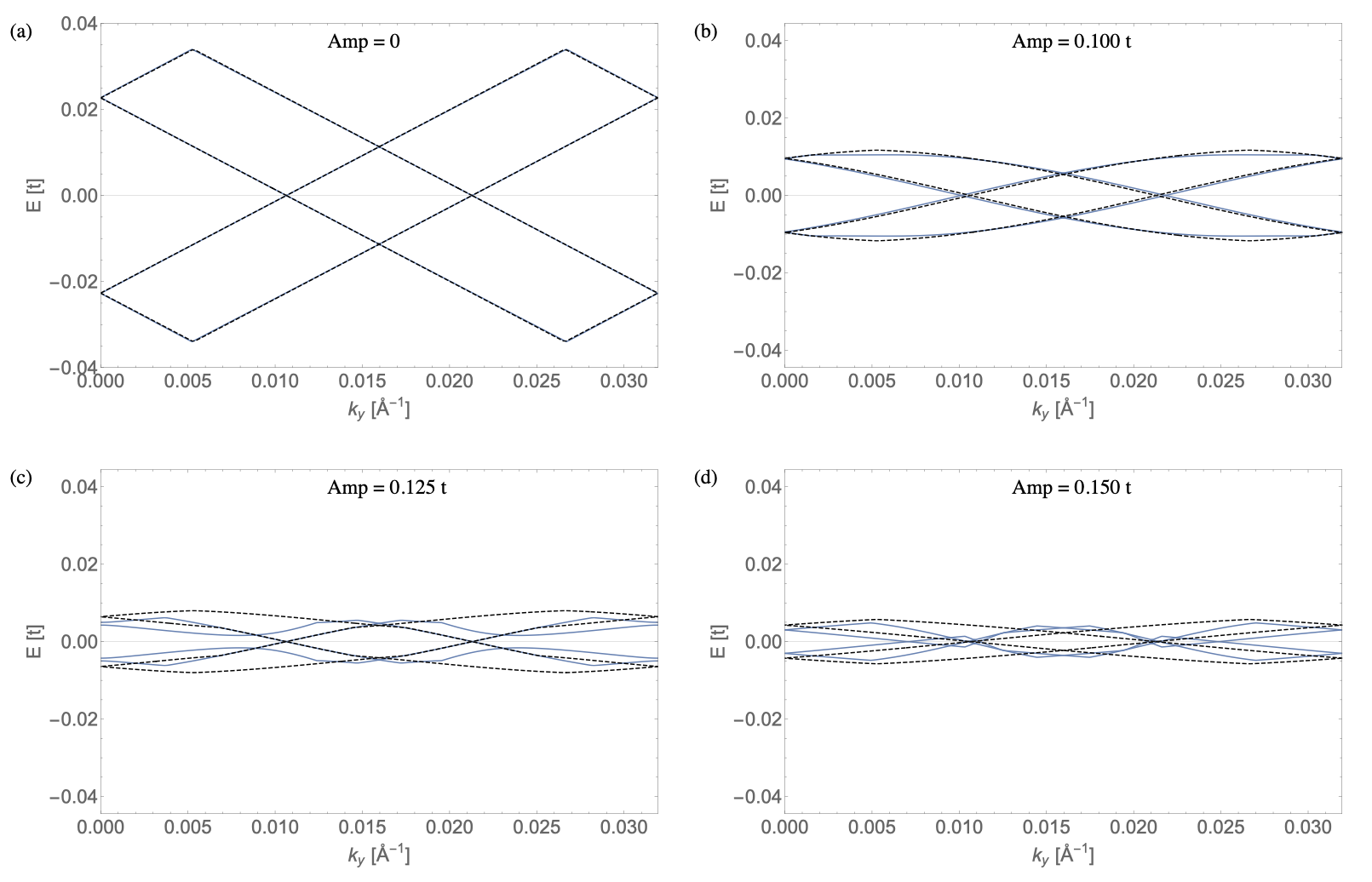}
\caption{\label{fig:bandCompare} Tight binding (solid blue) and continuum (dashed black) band structures for a checkerboard potential with periodicity of $204 \angstrom \times 197 \angstrom$, or $48 \times 80$ unit cells. (a) shows the case for amplitude of $0.000 t$, (b) shows $0.100t$, (c) shows $0.125t$, and (d) shows $0.150 t$.}
\end{figure*}

In this paper, we investigate the band structure of monolayer graphene gated by doubly periodic electrical potentials. In searching for band flattening, we consider three kinds of applied potentials: checkerboard, hexagonal, and Kagome lattice. Consistent with previous work~\cite{CheolHwan2008,Forsythe2018}, we find that the checkerboard potential can affect the Fermi velocity at the Dirac point, but we do not find strong examples of band flattening among energy bands near zero energy. Next, we consider two members of the wallpaper group 17, which has the same symmetries as those of both monolayer graphene as well as TBG~\cite{Zou2018}. For the case of a sinusoidal hexagonal potential, we are able to find gating voltages for which this potential induces strongly flattened bands. For the case of the Kagome potential, we find even stronger band flattening, with both the Fermi velocity and band curvature very close to zero in the vicinity of the Dirac point as shown in Figure~\ref{fig:plot3d}. We also investigate a potential with symmetries of wallpaper group 15, the results of which can be found in appendix~\ref{app:WG15}.

In our investigation we use both a tight-binding and a continuum model of gated graphene, as in references~\onlinecite{McCann_2013,CastroNeto2009}. These two models agree well within certain ranges of parameters. However, as the gating voltage is increased, the continuum model becomes less applicable as effects of the graphene lattice which cannot be described by physics near the Dirac cone alone become relevant. The tight-binding model considers these effects, but for lattices with larger periodicity we run into the limits of our computational resources. Further, the tight-binding model has its own pathologies associated with the comensurability of the applied potential lattice and the underlying graphene lattice that is necessary for numerical analysis, which are addressed in appendix~\ref{app:pathology}. An example of the divergence of the two models can be seen in Figure~\ref{fig:bandCompare}. In this figure, we plot the band structure computed using both models for a checkerboard potential and find that for low amplitudes, the models agree well. However, as the amplitude of the potential increases, the band structure obtained from each model differs more significantly. Therefore, in order to find the most robust band flattening scenario, we specifically look for band flattening that occurs in both the tight-binding and the continuum model at the same time.


Having an experimental knob for tuning band flatness, we propose an experimental investigation of whether band flattening is sufficient to reproduce the properties of MATBG with the use of only a single layer of graphene. This approach would help to experimentally disentangle whether flat bands are the fundamental characteristic that induces correlated insulator and superconducting phases in these systems or whether additional features~\cite{Andrei2020} such as fragile topology~\cite{Po2018,Po2019} are required.
	
In addition to the search for behavior similar to that of MATBG, our work points to the possibility for band engineering. Our models provide a predictive guide in determining the behavior of electronic band structure under different potentials. By modeling the results of the interplay of the applied potential and the intrinsic electronic properties of graphene, we make strides toward turning graphene into a metamaterial with fine control of velocity near the Dirac point~\cite{Forsythe2018,Shi2019}. 
We expect that continuous tuning of band flatness by electrostatic gating could be a powerful tool for probing fundamental physics of strongly correlated systems. Specifically, band flattening quenches the kinetic energy relative to the interaction energy, thus pushing graphene into the strongly interacting regime. Band structure engineering could also have potential applications in optics and electronics~\cite{Chaves2020}.

\section{Tight-Binding Model:}
Tight-binding calculations used in this paper consider the hopping of electrons between neighboring carbon atoms allowing us to derive the electronic band structure of the gated graphene lattice. Our tight-binding Hamiltonian consists of a hopping term that couples neighboring lattice sites and an on-site potential term that describes the action of an electrostatic gate. Here, we consider the nonmagnetic case and hence focus on only one spin species as we expect both to have an identical band structure. Therefore, the Hamiltonian is of the form
\begin{align}
    H=-t \Sigma_{i,j}(c_i^\dagger c_j + \text{h.c.})+\Sigma_i c_i^\dagger c_i V(\mathbf{r_i}),
\end{align}
where $i,j$ index the lattice sites of the honeycomb lattice, $c_i, c_i^\dagger$ are the annihilation and creation operators on site $i$, $t$ is the hopping term between lattice sites, and $V(\mathbf{r_i})$ is the applied potential at position $\mathbf{r_i}$, or the position of the site $i$.

However, in order to investigate this model in a computationally tractable way, we must apply some constraints to the choice of potential $V(\mathbf{r})$. The potential must be periodic in both spatial directions in such a way that the graphene lattice can be specified to have the same period in such directions, thus resulting in an extended unit cell that can be specified with finite computational resources. That is, the period of the potential in $x$ and $y$ directions must be an integer multiple of the size of a graphene unit cell. In the armchair configuration of the graphene unit cell as used in this simulation, the size of each cell is $3a \times \sqrt{3}a$, where $a$ is the distance between lattice sites. Only certain potentials at certain orientations relative to the graphene lattice are able to be made compatible with the integer unit cell requirement.

In particular, a square potential cannot be perfectly simulated by this tight-binding method due to the fact that it is not possible to find four points in a hexagonal lattice which form a square. Therefore, for our checkerboard potential, we use a rectangular potential by choosing the period in each direction to be an arbitrary integer multiple of the size of the graphene unit cell. Hexagonal lattices are possible as the graphene lattice itself is hexagonal in nature, however only certain angles of the potential relative to the graphene are possible. These commensurate angles are further classified into type-I and type-II, which are considered for Moiré lattices in twisted bilayer graphene by Zou and colleagues~\cite{Zou2018} and are extended here for hexagonal potentials applied to graphene. Type-I lattices are manifested in this simulation by those angles wherein the extended unit cell contains three vertical periods of the potential, whereas Type-II lattices are represented by extended unit cells containing a single period of the potential. Due to phenomena associated with band folding in the tight-binding model, type-I hexagonal lattices are not considered (see appendix~\ref{app:pathology}).

\section{Continuum Model}
We begin with the effective two-band low-energy Hamiltonian near the K points. In particular, we consider the Hamiltonian of the form
\begin{align}
    H=\begin{pmatrix}
-iv_0\boldsymbol{\sigma} \nabla & V(\mathbf{r}) \\
V(\mathbf{r}) & -iv_0\boldsymbol{\sigma} \nabla 
\end{pmatrix}
\end{align}
where $v_0$ represents the Fermi velocity in graphene, $\boldsymbol{\sigma}$ is the Pauli matrix vector, $\nabla=(\partial_x,\partial_y)$, and $V(\mathbf{r})$ represents the applied potential over the graphene lattice. We write the real-space wave function as in reference~\cite{Tarnopolsky2019} in terms of the momentum components, $a_{\sigma,ij}(\mathbf{k})$, as
\begin{align}
\psi_\mathbf{k}(\mathbf{r})=\Sigma_{i,j}
\begin{pmatrix}
a_{1,ij}(\mathbf{k}) \\
a_{2,ij}(\mathbf{k})
\end{pmatrix}
e^{i(\mathbf{K_{ij}}+\mathbf{k})\mathbf{r}},
\end{align}
where $\mathbf{K_{ij}}=i \mathbf{b_1} + j \mathbf{b_2}$ and $\mathbf{b_1},\mathbf{b_2}$ are the reciprocal lattice vectors of the extended unit cell of the periodic potential.


We wish to write the Hamiltonian in momentum space, and since the Hamiltonian is periodic in real space, the lattice momentum, which is set by the period of the potential, is a good quantum number. Therefore, the momentum space Hamiltonian is
\begin{widetext}
\label{H:scatt}
\begin{align}
    H=\begin{pmatrix}
0 & k_x-i k_y & F_{01} & 0 & \dots \\
k_x+i k_y & 0 & 0 & F_{01} & \dots \\
F_{10} & 0 & 0 & k_x+K_{10,x}-i( k_y + K_{10,y}) & \dots \\
0 & F_{10} & k_x+K_{10,x}+i(k_y+K_{10,y}) & 0 & \dots \\
\vdots & \vdots & \vdots & \vdots & \ddots
\end{pmatrix}
\end{align}
\end{widetext}
where the diagonal $2 \times 2$ blocks correspond to the kinetic energy shifted by lattice momentum $\mathbf{K}_{ij}$, while the off-diagonal $2 \times 2$ blocks at position $(i_1,j_1)(i_2,j_2)$ correspond to the Fourier transform of the potential energy with the coefficient $F_{(i_2-i_1)(j_2-j_1)}$ where the momentum transfer is $\mathbf{K}_{(i_2-i_1)(j_2-j_1)}$. When performing the numerics, we truncate the Hamiltonian matrix by enforcing $-n \leq i \leq n$ and $-n \leq j \leq n$, and choosing $n$ sufficiently large to ensure the convergence of the bands that we are interested in.

\begin{figure}
\includegraphics[width=\columnwidth]{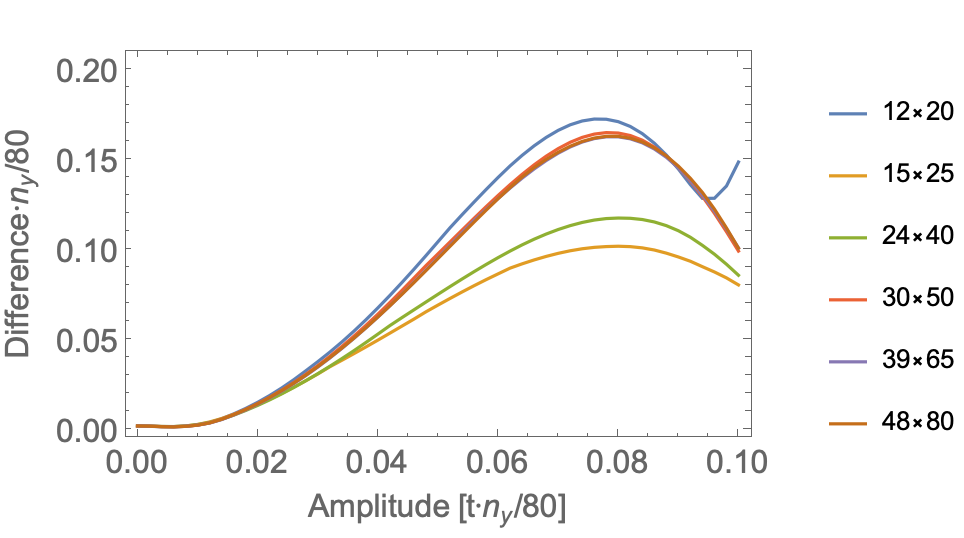}
\caption{\label{fig:modelCompare} Divergence of tight-binding and continuum models as amplitude of the checkerboard potential is increased. Each curve plots the difference between the two models for a different extended unit cell size. The differences and amplitudes here have been scaled by $n_y/80$, where $n_y$ is the number of graphene unit cells along the $y$ direction in the extended unit cell.
}
\end{figure}

The continuum model is only valid over a range of parameters for which effects far from the Dirac point are negligible, i.e. for momentum close to the Dirac cone center. As the amplitude of the applied potential increases, Eq.~\eqref{H:scatt} indicates that scattering between farther away $k$ points becomes more important, and the linear dispersion assumption of the continuum model breaks down. Consequently, the continuum model diverges from the tight-binding model and hence the true behavior of graphene. In particular, the model functions best for potentials with amplitudes less than $\approx 0.1 t$, where $t$ is the hopping term between lattice sites, and this term is given by $t=2\hbar v_0/3a$. In order to gauge the the deviation between the two models, we consider the normalized mean error between the central bands as computed by the two models along $k_y$, averaged over one Brillouin zone
\begin{align}
    \Delta=\frac{\int_{0}^{2 \pi/a} dk_y |\epsilon^{\text{TB}}_{k_x=0,k_y}-\epsilon^{\text{cont}}_{1, k_x=0,k_y}|}{\int_{0}^{2 \pi/a} dk_y \epsilon^{\text{TB}}_{1,k_x=0,k_y}}.
\end{align}
In Figure~\ref{fig:modelCompare}, we plot the difference $\Delta$ scaled by $n_y/80$ as a function of the amplitude of the checkerboard potential also scaled by $n_y/80$ for various extended unit cell sizes $n_x \times n_y$. We observe reasonable collapse of the numerical data, indicating that regardless of extended unit cell size, the models diverge at roughly the same rate as amplitude is increased.



\section{Results}
We find instances of band flattening which show promise of being experimentally reproducible, with verification of band flattening in both tight-binding and continuum model calculations and for various gating potentials. We use our continuum model to numerically compute band structure and identify band flattening near high-symmetry points ($k_x=0,k_y=0$) over regimes spanning a range of amplitudes. We then compute the second derivatives of these bands in order to identify a gating strength with optimal flattening. The tight-binding model provides an additional check on these calculations to ensure we are operating within the applicable regime of parameters of the continuum model. We consider here the results from a checkerboard potential, a sinusoidal hexagonal potential, and a Kagome potential.

In addition to these useful regimes of band flattening, the tight-binding model implies instances of band flattening which occur at very precisely tuned values of applied amplitude, in particular at points wherein the observed Dirac point splits into further Dirac points. However, we note that these cases of band flattening are likely difficult to reproduce experimentally as variations of the extended unit cell size by a single graphene cell will destabilize the flat point. This indicates that experimental realization of checkerboard-based band flattening requires sub-nanometer control of the applied potential as well as fine tuning of the potential amplitude. While these phenomena exist in principle, they do not provide a strong basis for demonstrating flat bands in a lab. See appendix~\ref{app:pathology} for more.

\begin{figure}
\includegraphics[width=\columnwidth]{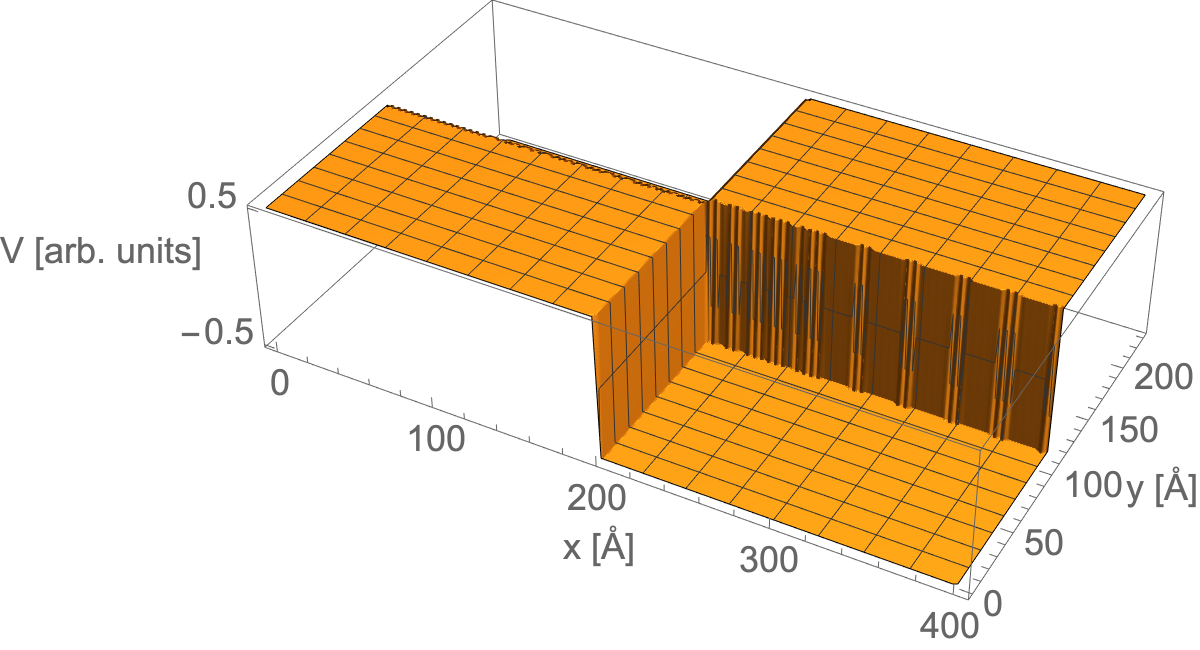}
\caption{\label{fig:potentialChecker} Single extended unit cell of a checkerboard potential with different periodicity in each direction. This potential is used to produce the results in Figure~\ref{fig:derivs}, parts (a) and (d).}
\end{figure}

\begin{figure*}
\includegraphics[width=\textwidth]{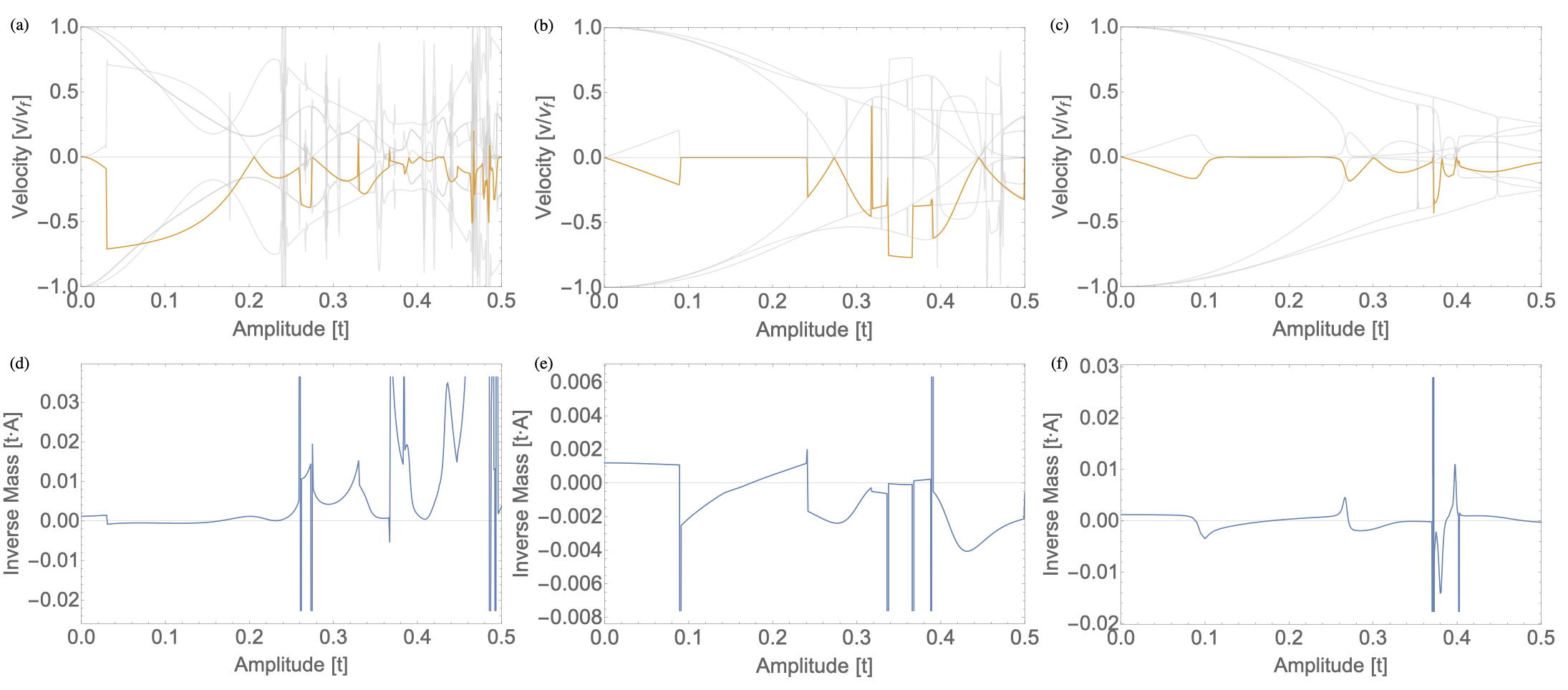}
\caption{\label{fig:derivs} Numerically computed derivatives of the energy bands from the continuum model near Brillouin zone center versus applied potential amplitude. (a),(b), and (c) show a selection of first derivatives for energy bands near the zero energy for the checkerboard, sinusoidal hexagonal, and Kagome potentials, respectively. In each, one energy band is highlighted in orange. (d), (e), and (f) show the second derivative of the energy band whose first derivative is highlighted in (a),(b), and(c), respectively. The amplitude is measured in units of $t$, the velocity is measured relative to the Fermi velocity of electrons in graphene, $v_f$, and the inverse mass is measured in units of $t \cdot A$, where $A$ is the extended unit cell area.}
\end{figure*}


\subsection{Checkerboard Potential}
The checkerboard potential, a member of wallpaper group 11 (p4mm), pictured in Figure~\ref{fig:potentialChecker}, is used as an extension of the 1D case of a square potential as investigated by Li and colleagues~\cite{Li2021}, as it is composed of square waves along the $x$ and $y$ directions independently. It is of the form
\begin{align}
    V(x,y)=V_0 \left(\text{IF}((x,y) \in \text{High},0.5,-0.5)\right), \nonumber
\end{align}
where $V_0$ is the amplitude of the potential, the IF function returns the second argument if the first is true and the third argument otherwise, and the set ``High'' refers to the set of points $(x,y)$ that lie within a tile in the checkerboard with the higher potential.
We apply this potential such that the periodicity in each direction is different because of the required periodicity imposed by the graphene lattice. That is, the tight-binding model requires that an extended unit cell be specified which is doubly periodic, and therefore takes into account the periodicity of both the graphene and the applied potential in both directions. However, the hexagonal lattice of the graphene is not commensurate with a square lattice, and it is not possible to choose 4 graphene sites in a honeycomb lattice such that they form a square. Therefore, the tight-binding model can only provide results for a rectangular potential, where the square potential is slightly stretched to fit the chosen dimensions of the extended unit cell in the model.

In Figure~\ref{fig:bandCompare}, we plot the electronic band structures computed by the tight binding model and the continuum model at select amplitudes of the checkerboard potential. We observe that both models show band flattening over the range of amplitudes chosen, as the energy levels of the plotted bands decrease toward zero. Additionally, as amplitude is increased, the models begin to disagree.

In Figure~\ref{fig:derivs}, we see plotted the derivatives of several energy bands near zero energy over a range of amplitudes for a checkerboard potential with periodicity of 94 graphene unit cells in each direction ($400\angstrom \times 231 \angstrom$). This checkerboard potential case demonstrates the least band flattening compared to other potentials investigated in this paper, with few regions of amplitude over which any band near the central bands are flattened. That is, band flattening to a velocity of zero occurs only at finely tuned values of gating strength rather than over larger amplitude ranges. This is the case for amplitudes up to half of the coupling energy and for bands near zero energy. Because of this, we do not identify any applied amplitudes wherein the first and second derivatives of the energy bands near the Dirac points are simultaneously nearly zero. Therefore, we do not find the checkerboard potential to be useful in attaining flat bands in graphene for relevant applied potential strengths.


\begin{figure}
\includegraphics[width=\columnwidth]{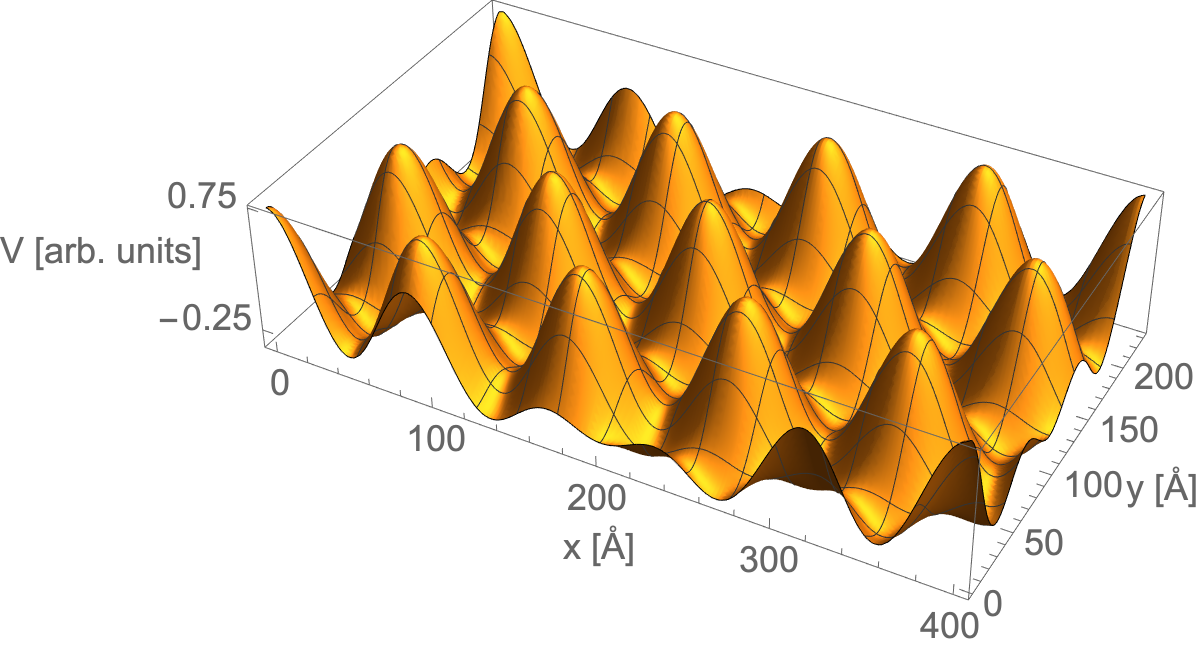}
\caption{\label{fig:potentialHex} Single extended unit cell of the sinusoidal hexagonal potential with rotation of approximately 10.9$\degree$ relative to the graphene lattice. This potential is used to produce the results in Figure~\ref{fig:derivs}, parts (b) and (e).}
\end{figure}

\subsection{Sinusoidal Hexagonal Potential}
This potential, which is a member of wallpaper group 17 (p6mm) and which can be expressed without rotation as 
\begin{align}
    V(x,y)=\frac{V_0}{4} \left(\cos(\frac{2y}{\sqrt 3})+\cos(\frac{y}{\sqrt 3}+x)+\cos(\frac{y}{\sqrt 3}-x)\right), \nonumber
\end{align}
where $V_0$ is the amplitude of the potential, is commensurate with the graphene lattice due to its hexagonal nature, and thus is replicable in the tight-binding model. For example, see Figure~\ref{fig:potentialHex}, which shows a sinusoidal hexagonal potential that is rotated by 10.9$\degree$ so as to be commensurate with the underlying graphene lattice. In this example, which shows the same potential as used for the simulations in Figure~\ref{fig:derivs}, the extended unit cell is composed of $94 \times 94$ graphene unit cells, resulting in a size of $400\angstrom \times 231 \angstrom$. However, the distance between adjacent potential maxima is only $87 \angstrom$. The extended unit cell size is chosen so as to maintain the required double-periodicity, accounting for both the potential and the graphene lattice.

For this potential, we find that certain bands are flattened for a range of applied gating voltages, as shown in Figure~\ref{fig:derivs}. We are able to identify at least one point where both the first and second derivative of one band are simultaneously nearly zero, as highlighted in parts (b) and (e) of the figure. We also identify further points at which the Fermi velocity becomes small for finely tuned values of the applied gating voltage. Therefore, a hexagonal potential is a reasonable choice for exploring phenomena expected of flat bands. In particular, for $400\angstrom \times 231 \angstrom$ extended unit cell size the amplitude of the applied potential should be $0.45t$. If we scale the size of the extended unit cell, while keeping its aspect ratio fixed (i.e. $L_x/L_y=400/231$), then the amplitude of the applied potential needed to obtain flat bands scales as $(400 \angstrom/L_x) * (0.45 t)$.

\begin{figure}
\includegraphics[width=\columnwidth]{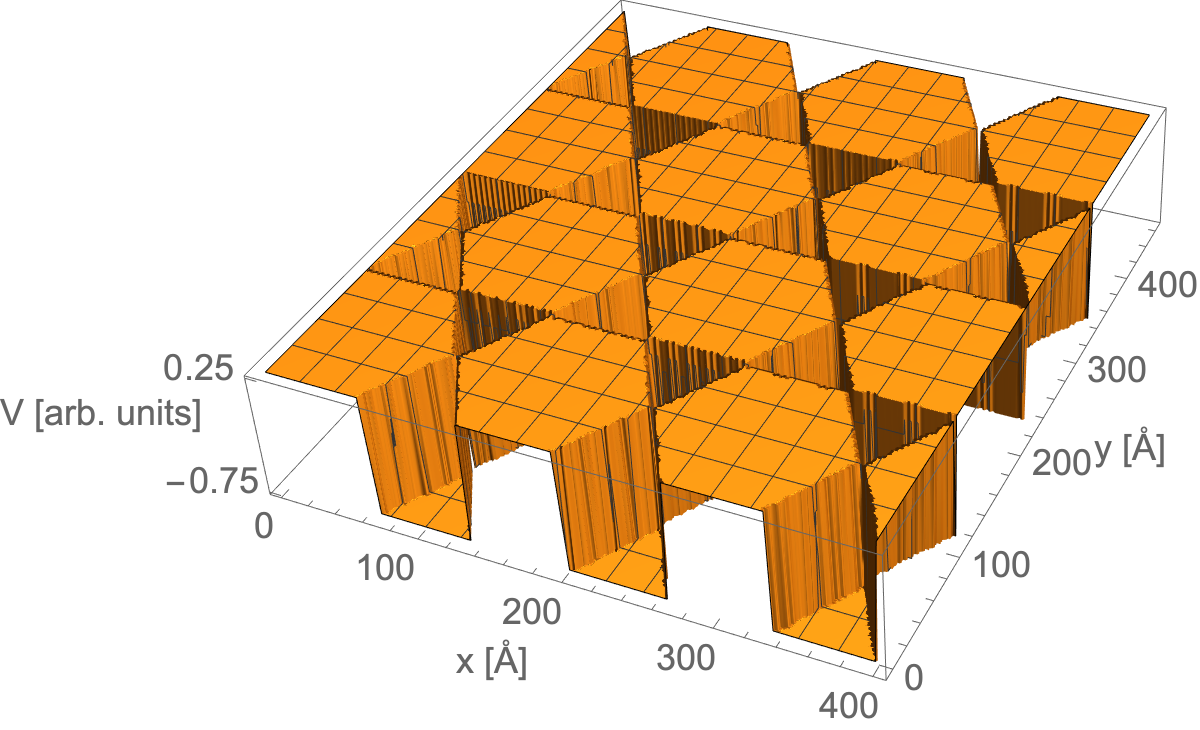}
\caption{\label{fig:potentialWG17} Kagome potential used in this simulation. Here, we plot two copies of the extended unit cell in the $y$ direction, and one copy in the $x$ direction. This potential is used to produce the results in Figure~\ref{fig:derivs}, parts (c) and (f).}
\end{figure}


\subsection{Kagome Potential}
The Kagome potential has a hexagonal lattice and can therefore be represented with tight-binding calculations in a way which is commensurate with the underlying graphene lattice. This potential is a member of wallpaper group 17 (p6mm), which represents the same symmetries as twisted bilayer graphene. In this paper, we consider a uniform 2-coloring of the Kagome lattice as pictured in Figure~\ref{fig:potentialWG17}. In particular, the potential is of the form
\begin{align}
    V(x,y)=V_0 \left(\text{IF}((x,y) \in \text{Hexagon},0.25,-0.75)\right), \nonumber
\end{align}
where $V_0$ is the amplitude of the potential, the IF function returns the second argument if the first is true and the third argument otherwise, and the set ``Hexagon'' refers to the set of points $(x,y)$ that lie within a hexagonal tile in the Kagome potential. The hexagons are assigned a potential of 0.25 and the triangles a potential of -0.75 in order to set the average potential to be 0.

We show in part (c) of Figure~\ref{fig:derivs} the first derivatives of bands calculated using the continuum model for this potential. We observe significant ranges of applied voltage where bands near zero energy have nearly zero velocity. At the same time, we are able to identify points where the second derivative of a band is zero and coincides with zero velocity, as highlighted in part (f) of the figure. The wide range of amplitudes for which the Kagome potential exhibits flat bands indicates that this potential is the most promising for producing the desired flat band effects in practice, particularly near an applied potential amplitude of $0.177t$. Although it is tempting to ascribe the presence of these flat bands to the effects of a Kagome lattice of sites as found by Bergman and colleagues in~\onlinecite{Bergman2008}, we do not expect this is the case as we observe a pair of flat bands in our region of interest rather than a single band as predicted by Bergman's Kagome lattice model.

\section{Discussion}
In summary, we have shown that it is possible to induce band flattening in gated graphene through the application of doubly periodic potentials. We have also identified a range of potential amplitudes over which this is possible, and our results indicate that the Kagome potential shows the most robust band flattening. In particular, a Kagome potential with spacing between hexagon centers of, for instance, 11.6~nm requires a gating potential of  5.5~eV to get flat bands, while one with 50~nm spacing requires 1.3~eV (where we use $v_f \approx 10^6$ m/s \cite{CastroNeto2009}). Electronically reprogrammable complex oxide heterostructures, in particular LaAlO$_3$/SrTiO$_3$ interfaces~\cite{cen2008}, can potentially be used to produce these doubly periodic potentials. Graphene/LaAlO$_3$/SrTiO$_3$ heterostructures can be programmed using conductive atomic force microscope (c-AFM) lithography~\cite{huang2015, jnawali2018}; however, this technique is not amenable to van der Waals stacks containing hexagonal boron nitride. A more recently developed approach, which uses ultra-low-voltage electron beam lithography (ULV-EBL) to achieve nanoscale control of the metal insulator transition at buried LaAlO$_3$/SrTiO$_3$ interfaces~\cite{yang2020}, can realize sub-10-nm resolution conductive nanostructures that are reprogrammable, and thus provide a pathway toward a more general approach to analog quantum simulation. The theoretical results described here offer guidance and insight into the families of doubly periodic potentials that are likely to induce strong electron-electron interactions due to ultraflat electronic dispersion.

\section{Acknowledgements}
This work was supported by NSF PHY-1913034. RR and JL also acknowledge support from the DOE-QIS program.

\appendix	
\section{Pathologies of the tight-binding model}
\label{app:pathology}

\begin{figure*}
\includegraphics[width=\textwidth]{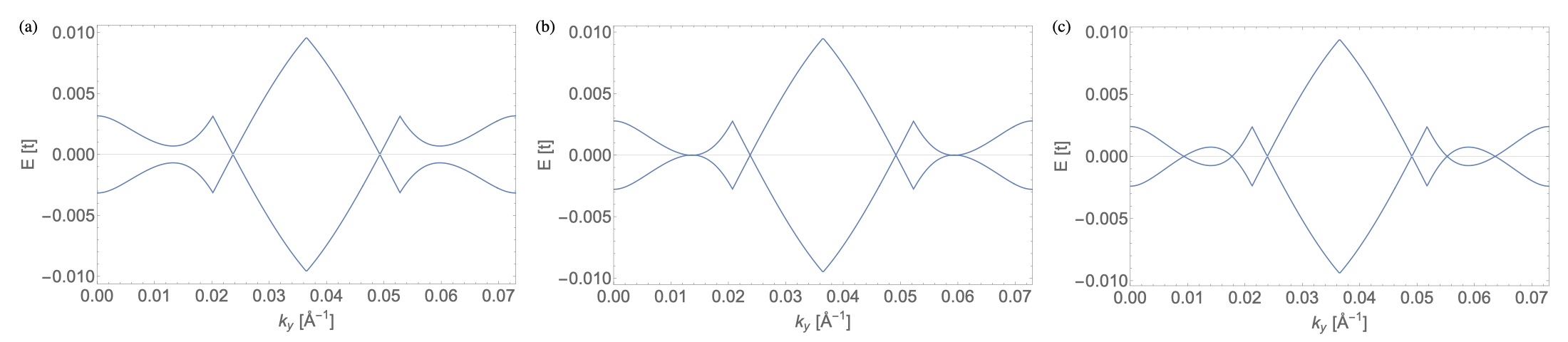}
\caption{\label{fig:pathFlat} Band structure for checkerboard potential with extended unit cell size $85 \angstrom \times 86 \angstrom$, or $20 \times 35$ graphene unit cells. In (a), the potential has an amplitude of $0.2850t$, and the central bands approach each other with a point of minimum separation at $k_y=$ and $k_y=$. In (b), at an amplitude of $0.2874t$, the two bands touch and form a flat region at $k_y=$ and $k_y=$. In (c), with a potential amplitude of $0.2900t$, the flat region has split into two distinct Dirac points.}
\end{figure*}

\begin{figure*}
\includegraphics[width=\textwidth]{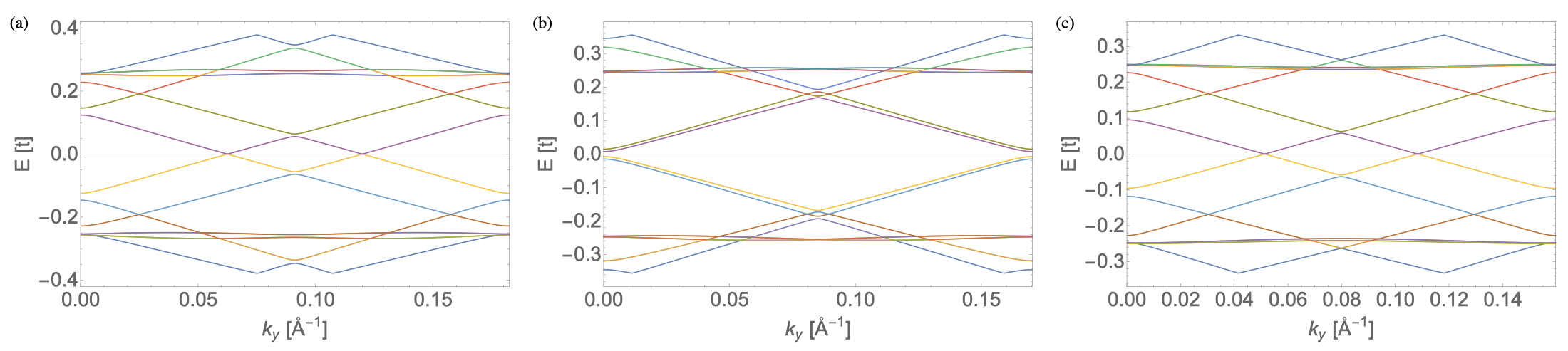}
\caption{\label{fig:pathology1} Example of band folding effects at certain unit cell sizes for a 1D square wave which is periodic along the $y$ direction. In (a), we have a $10 \times 14$ unit cell potential, in (b), a $10 \times 15$ unit cell potential, and in (c), a $10 \times 16$ unit cell potential. All three have an applied potential of amplitude $0.3 t$. The band folding in (b) produces an electronic band structure wherein the two Dirac points in the Brillouin zone are mapped to the same point, $(k_x,k_y)=(0,0)$, so that the Dirac point is split. Due to the precise unit cell size needed to realize this, we expect that this effect is difficult to reproduce in a laboratory setting.}
\end{figure*}

Due to the discrete nature of the tight-binding model and its underlying lattice, we note that the model predicts various phenomena which one would not reasonably expect to be replicable in a laboratory setting. There are two important classes of such spurious phenomena. The first is band flattening which occurs only at precise, finely tuned values of the applied potential. We observe this at transitionary points where the band touching point metamorphoses into a pair of Dirac points as the amplitude is tuned. The second occurs when the number of simulated graphene cells in the vertical direction, i.e. the direction in which the graphene unit cell has length $\sqrt{3}a$, is a multiple of 3. In this case, band folding will perfectly map the two Dirac cones in a unit cell to the same point in the Brillouin zone associated with the potential, resulting in an electronic band structure that is very sensitive to the commensurability of the graphene and the electrostatic potential lattices.

Here, we note that our tight-binding model can demonstrate band flattening at finely tuned values of the potential which, while plausible in theory, are unlikely to be accessible via experiment. These occur as the amplitude is tuned such that two bands approach each other until they touch at a point, and then split into two Dirac points, as shown in Figure~\ref{fig:pathFlat}. This process can also happen in reverse order with amplitude tuning. Additionally, as the periodicity of the applied potential is varied by a single graphene unit cell (on the order of 5$\angstrom$), the necessary applied amplitude to see band flattening varies significantly. Therefore, experimental realization of these modes of band flattening would require precise control of both the amplitude of the applied potential and the length scale over which the potential is applied. As demonstrated by Barbier and colleagues, a continuum model of graphene can predict the emergence of extra Dirac points at certain applied potential strengths in the case of a one-dimensional potential~\cite{Park2008a,Park2009,Brey2009,Barbier2010,Dubey2013}. However, these additional Dirac points are not analogous to the anomalous Dirac points we see in our tight-binding model. Rather, these Dirac points, which are predicted by a continuum model, are stable over a range of applied potential amplitudes. The anomalous Dirac points discussed here are predicted only by our tight-binding model for two-dimensional potentials and are not stable under changes to the applied potential.

Band folding effects are responsible for the second type of anomaly of the tight-binding model discussed here. This effect, which can be seen in Figure~\ref{fig:pathology1}, occurs when the size of the chosen extended unit cell is such that the two distinct Dirac points are mapped to the same point in $k$-space. In particular, this occurs when the number of graphene unit cells in the super cell in the $y$ direction is divisible by 3. This is because for a single graphene unit cell, Dirac points occur along the $k_y$ direction at exactly thirds of the size of the Brillouin zone. Therefore, when the number of unit cells is such that the Brillouin zone is reduced by a factor divisible by 3, both of the Dirac points will all be mapped to the zone center. In practice, these effects are unlikely to be detectable experimentally as they require the precise alignment of the periodic potential to the graphene lattice with precision of better than one graphene lattice spacing. Consequently, we consider only cases where the period of the applied potential in the vertical direction is not precisely a multiple of three graphene unit cells.

This second effect prevents useful simulation of type-I hexagonal lattices, i.e. those where the chosen rectangle of lattice sites is such that the number of vertical graphene unit cells is a multiple of three. This limits the set of angles at which we can simulate hexagonal lattices to those for which a type-II lattice can be constructed.

\appendix	
\section{Wallpaper group 15}
\label{app:WG15}
\begin{figure}
\includegraphics[width=\columnwidth]{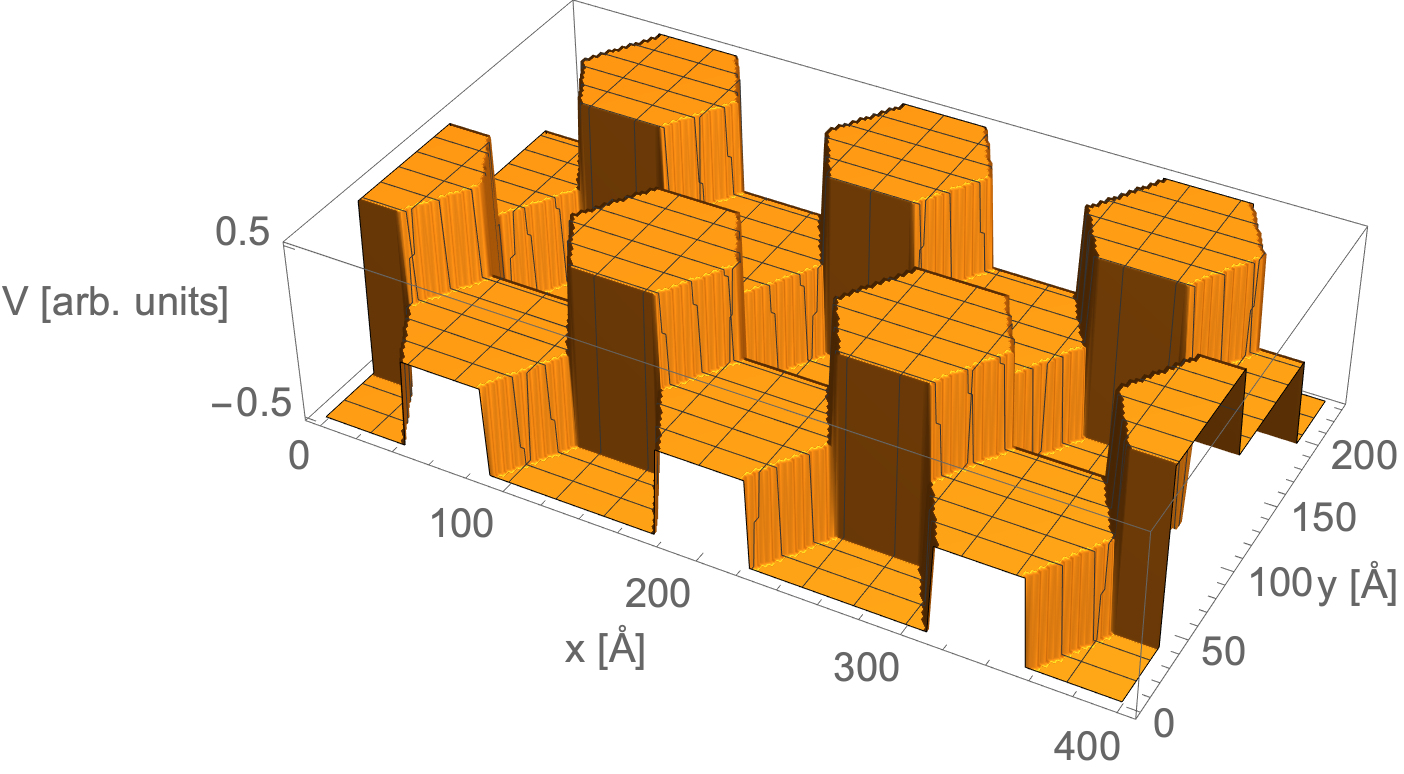}
\caption{\label{fig:potentialCatan} Single extended unit cell of a wallpaper group 15 potential. This potential is used to produce the results in Figure~\ref{fig:wg15derivs}.}
\end{figure}

\begin{figure}
\includegraphics[width=\columnwidth]{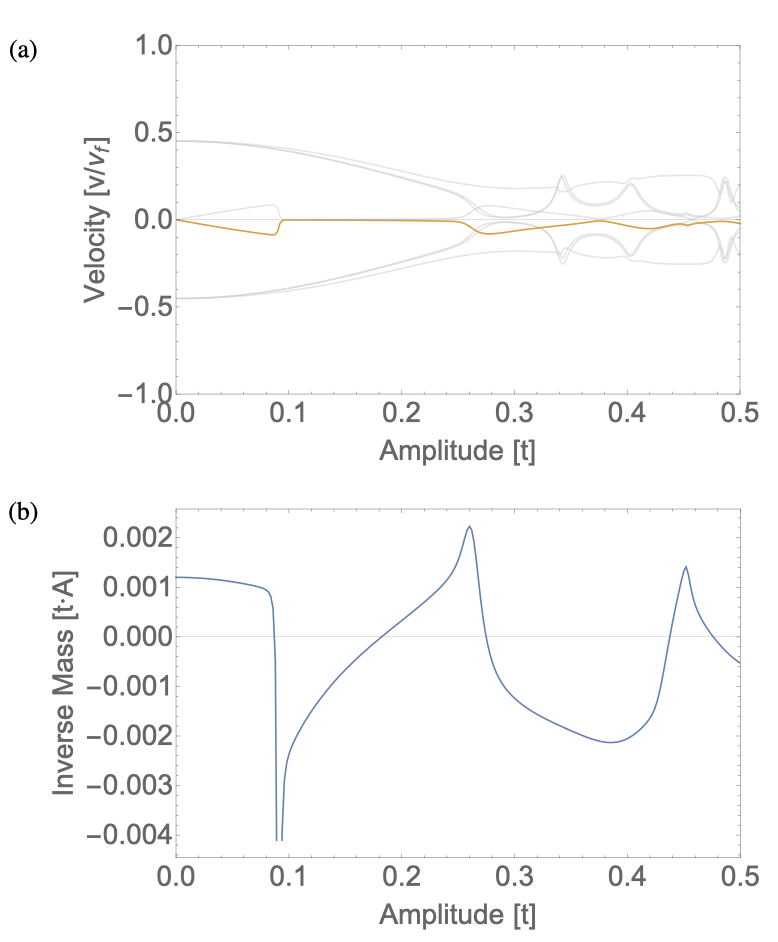}
\caption{\label{fig:wg15derivs} Numerically computed derivatives of the energy bands from the continuum model near Brillouin zone center versus applied potential amplitude for the wallpaper group 15 potential. (a) shows a selection of first derivatives for energy bands near zero energy. One energy band with a region of flatness over a range of amplitudes is highlighted in orange. (b) shows the second derivative of the highlighted energy band.}
\end{figure}

In addition to the three potentials outlined above, we consider an additional gating potential which is a member of wallpaper group 15 (p3m1). This potential consists of a 3-coloring of a hexagonal lattice, with a potential expressed by
\begin{align}
    V(x,y)=V_0
    \begin{cases} 
      0.5 & (x,y)\in \text{Hex1} \\
      0 & (x,y)\in \text{Hex2} \\
      -0.5 & (x,y)\in \text{Hex3} 
   \end{cases}
, \nonumber
\end{align}
where ``Hex1'', ``Hex2'', and ``Hex3'' are the sets of points such that each set represents hexagons at one of the assigned potential levels. This regular coloring of a hexagonal grid is pictured in Figure~\ref{fig:potentialCatan}.
We find that this potential results in band flattening similar to that which is seen with the Kagome potential. These results, shown in Figure~\ref{fig:wg15derivs}, demonstrate that we are able to find bands which exhibit flatness for a range of gating amplitudes for the wallpaper group 15 potential.

\bibliography{apssamp}

\providecommand{\noopsort}[1]{}\providecommand{\singleletter}[1]{#1}%
\begin{thebibliography}{38}%
\makeatletter
\providecommand \@ifxundefined [1]{%
 \@ifx{#1\undefined}
}%
\providecommand \@ifnum [1]{%
 \ifnum #1\expandafter \@firstoftwo
 \else \expandafter \@secondoftwo
 \fi
}%
\providecommand \@ifx [1]{%
 \ifx #1\expandafter \@firstoftwo
 \else \expandafter \@secondoftwo
 \fi
}%
\providecommand \natexlab [1]{#1}%
\providecommand \enquote  [1]{``#1''}%
\providecommand \bibnamefont  [1]{#1}%
\providecommand \bibfnamefont [1]{#1}%
\providecommand \citenamefont [1]{#1}%
\providecommand \href@noop [0]{\@secondoftwo}%
\providecommand \href [0]{\begingroup \@sanitize@url \@href}%
\providecommand \@href[1]{\@@startlink{#1}\@@href}%
\providecommand \@@href[1]{\endgroup#1\@@endlink}%
\providecommand \@sanitize@url [0]{\catcode `\\12\catcode `\$12\catcode
  `\&12\catcode `\#12\catcode `\^12\catcode `\_12\catcode `\%12\relax}%
\providecommand \@@startlink[1]{}%
\providecommand \@@endlink[0]{}%
\providecommand \url  [0]{\begingroup\@sanitize@url \@url }%
\providecommand \@url [1]{\endgroup\@href {#1}{\urlprefix }}%
\providecommand \urlprefix  [0]{URL }%
\providecommand \Eprint [0]{\href }%
\providecommand \doibase [0]{https://doi.org/}%
\providecommand \selectlanguage [0]{\@gobble}%
\providecommand \bibinfo  [0]{\@secondoftwo}%
\providecommand \bibfield  [0]{\@secondoftwo}%
\providecommand \translation [1]{[#1]}%
\providecommand \BibitemOpen [0]{}%
\providecommand \bibitemStop [0]{}%
\providecommand \bibitemNoStop [0]{.\EOS\space}%
\providecommand \EOS [0]{\spacefactor3000\relax}%
\providecommand \BibitemShut  [1]{\csname bibitem#1\endcsname}%
\let\auto@bib@innerbib\@empty
\bibitem [{\citenamefont {Wallace}(1947)}]{Wallace1947}%
  \BibitemOpen
  \bibfield  {author} {\bibinfo {author} {\bibfnamefont {P.~R.}\ \bibnamefont
  {Wallace}},\ }\bibfield  {title} {\bibinfo {title} {The band theory of
  graphite},\ }\href {https://doi.org/10.1103/physrev.71.622} {\bibfield
  {journal} {\bibinfo  {journal} {Physical Review}\ }\textbf {\bibinfo {volume}
  {71}},\ \bibinfo {pages} {622} (\bibinfo {year} {1947})}\BibitemShut
  {NoStop}%
\bibitem [{\citenamefont {Novoselov}\ \emph {et~al.}(2004)\citenamefont
  {Novoselov}, \citenamefont {Geim}, \citenamefont {Morozov}, \citenamefont
  {Jiang}, \citenamefont {Zhang}, \citenamefont {Dubonos}, \citenamefont
  {Grigorieva},\ and\ \citenamefont {Firsov}}]{Novoselov2004}%
  \BibitemOpen
  \bibfield  {author} {\bibinfo {author} {\bibfnamefont {K.~S.}\ \bibnamefont
  {Novoselov}}, \bibinfo {author} {\bibfnamefont {A.~K.}\ \bibnamefont {Geim}},
  \bibinfo {author} {\bibfnamefont {S.~V.}\ \bibnamefont {Morozov}}, \bibinfo
  {author} {\bibfnamefont {D.}~\bibnamefont {Jiang}}, \bibinfo {author}
  {\bibfnamefont {Y.}~\bibnamefont {Zhang}}, \bibinfo {author} {\bibfnamefont
  {S.~V.}\ \bibnamefont {Dubonos}}, \bibinfo {author} {\bibfnamefont {I.~V.}\
  \bibnamefont {Grigorieva}},\ and\ \bibinfo {author} {\bibfnamefont {A.~A.}\
  \bibnamefont {Firsov}},\ }\bibfield  {title} {\bibinfo {title} {Electric
  field effect in atomically thin carbon films},\ }\href
  {https://doi.org/10.1126/science.1102896} {\bibfield  {journal} {\bibinfo
  {journal} {Science}\ }\textbf {\bibinfo {volume} {306}},\ \bibinfo {pages}
  {666} (\bibinfo {year} {2004})},\ \Eprint
  {https://arxiv.org/abs/https://www.science.org/doi/pdf/10.1126/science.1102896}
  {https://www.science.org/doi/pdf/10.1126/science.1102896} \BibitemShut
  {NoStop}%
\bibitem [{\citenamefont {Novoselov}\ \emph
  {et~al.}(2005{\natexlab{a}})\citenamefont {Novoselov}, \citenamefont {Geim},
  \citenamefont {Morozov}, \citenamefont {Jiang}, \citenamefont {Katsnelson},
  \citenamefont {Grigorieva}, \citenamefont {Dubonos},\ and\ \citenamefont
  {Firsov}}]{Novoselov2005}%
  \BibitemOpen
  \bibfield  {author} {\bibinfo {author} {\bibfnamefont {K.~S.}\ \bibnamefont
  {Novoselov}}, \bibinfo {author} {\bibfnamefont {A.~K.}\ \bibnamefont {Geim}},
  \bibinfo {author} {\bibfnamefont {S.~V.}\ \bibnamefont {Morozov}}, \bibinfo
  {author} {\bibfnamefont {D.}~\bibnamefont {Jiang}}, \bibinfo {author}
  {\bibfnamefont {M.~I.}\ \bibnamefont {Katsnelson}}, \bibinfo {author}
  {\bibfnamefont {I.~V.}\ \bibnamefont {Grigorieva}}, \bibinfo {author}
  {\bibfnamefont {S.~V.}\ \bibnamefont {Dubonos}},\ and\ \bibinfo {author}
  {\bibfnamefont {A.~A.}\ \bibnamefont {Firsov}},\ }\bibfield  {title}
  {\bibinfo {title} {Two-dimensional gas of massless dirac fermions in
  graphene},\ }\href {https://doi.org/10.1038/nature04233} {\bibfield
  {journal} {\bibinfo  {journal} {Nature}\ }\textbf {\bibinfo {volume} {438}},\
  \bibinfo {pages} {197} (\bibinfo {year} {2005}{\natexlab{a}})}\BibitemShut
  {NoStop}%
\bibitem [{\citenamefont {Novoselov}\ \emph
  {et~al.}(2005{\natexlab{b}})\citenamefont {Novoselov}, \citenamefont {Jiang},
  \citenamefont {Schedin}, \citenamefont {Booth}, \citenamefont {Khotkevich},
  \citenamefont {Morozov},\ and\ \citenamefont {Geim}}]{Novoselov2005a}%
  \BibitemOpen
  \bibfield  {author} {\bibinfo {author} {\bibfnamefont {K.~S.}\ \bibnamefont
  {Novoselov}}, \bibinfo {author} {\bibfnamefont {D.}~\bibnamefont {Jiang}},
  \bibinfo {author} {\bibfnamefont {F.}~\bibnamefont {Schedin}}, \bibinfo
  {author} {\bibfnamefont {T.~J.}\ \bibnamefont {Booth}}, \bibinfo {author}
  {\bibfnamefont {V.~V.}\ \bibnamefont {Khotkevich}}, \bibinfo {author}
  {\bibfnamefont {S.~V.}\ \bibnamefont {Morozov}},\ and\ \bibinfo {author}
  {\bibfnamefont {A.~K.}\ \bibnamefont {Geim}},\ }\bibfield  {title} {\bibinfo
  {title} {Two-dimensional atomic crystals},\ }\href
  {https://doi.org/10.1073/pnas.0502848102} {\bibfield  {journal} {\bibinfo
  {journal} {Proceedings of the National Academy of Sciences}\ }\textbf
  {\bibinfo {volume} {102}},\ \bibinfo {pages} {10451} (\bibinfo {year}
  {2005}{\natexlab{b}})}\BibitemShut {NoStop}%
\bibitem [{\citenamefont {Geim}\ and\ \citenamefont
  {Novoselov}(2007)}]{Geim2007}%
  \BibitemOpen
  \bibfield  {author} {\bibinfo {author} {\bibfnamefont {A.~K.}\ \bibnamefont
  {Geim}}\ and\ \bibinfo {author} {\bibfnamefont {K.~S.}\ \bibnamefont
  {Novoselov}},\ }\bibfield  {title} {\bibinfo {title} {The rise of graphene},\
  }\href {https://doi.org/10.1038/nmat1849} {\bibfield  {journal} {\bibinfo
  {journal} {Nature Materials}\ }\textbf {\bibinfo {volume} {6}},\ \bibinfo
  {pages} {183} (\bibinfo {year} {2007})}\BibitemShut {NoStop}%
\bibitem [{\citenamefont {dos Santos}\ \emph {et~al.}(2007)\citenamefont {dos
  Santos}, \citenamefont {Peres},\ and\ \citenamefont
  {Neto}}]{LopesdosSantos2007}%
  \BibitemOpen
  \bibfield  {author} {\bibinfo {author} {\bibfnamefont {J.~M. B.~L.}\
  \bibnamefont {dos Santos}}, \bibinfo {author} {\bibfnamefont {N.~M.~R.}\
  \bibnamefont {Peres}},\ and\ \bibinfo {author} {\bibfnamefont {A.~H.~C.}\
  \bibnamefont {Neto}},\ }\bibfield  {title} {\bibinfo {title} {Graphene
  bilayer with a twist: Electronic structure},\ }\bibfield  {journal} {\bibinfo
   {journal} {Physical Review Letters}\ }\textbf {\bibinfo {volume} {99}},\
  \href {https://doi.org/10.1103/physrevlett.99.256802}
  {10.1103/physrevlett.99.256802} (\bibinfo {year} {2007})\BibitemShut
  {NoStop}%
\bibitem [{\citenamefont {Morell}\ \emph {et~al.}(2010)\citenamefont {Morell},
  \citenamefont {Correa}, \citenamefont {Vargas}, \citenamefont {Pacheco},\
  and\ \citenamefont {Barticevic}}]{SurezMorell2010}%
  \BibitemOpen
  \bibfield  {author} {\bibinfo {author} {\bibfnamefont {E.~S.}\ \bibnamefont
  {Morell}}, \bibinfo {author} {\bibfnamefont {J.~D.}\ \bibnamefont {Correa}},
  \bibinfo {author} {\bibfnamefont {P.}~\bibnamefont {Vargas}}, \bibinfo
  {author} {\bibfnamefont {M.}~\bibnamefont {Pacheco}},\ and\ \bibinfo {author}
  {\bibfnamefont {Z.}~\bibnamefont {Barticevic}},\ }\bibfield  {title}
  {\bibinfo {title} {Flat bands in slightly twisted bilayer graphene:
  Tight-binding calculations},\ }\bibfield  {journal} {\bibinfo  {journal}
  {Physical Review B}\ }\textbf {\bibinfo {volume} {82}},\ \href
  {https://doi.org/10.1103/physrevb.82.121407} {10.1103/physrevb.82.121407}
  (\bibinfo {year} {2010})\BibitemShut {NoStop}%
\bibitem [{\citenamefont {Bistritzer}\ and\ \citenamefont
  {MacDonald}(2011)}]{Bistritzer2011}%
  \BibitemOpen
  \bibfield  {author} {\bibinfo {author} {\bibfnamefont {R.}~\bibnamefont
  {Bistritzer}}\ and\ \bibinfo {author} {\bibfnamefont {A.~H.}\ \bibnamefont
  {MacDonald}},\ }\bibfield  {title} {\bibinfo {title} {Moire bands in twisted
  double-layer graphene},\ }\href {https://doi.org/10.1073/pnas.1108174108}
  {\bibfield  {journal} {\bibinfo  {journal} {Proceedings of the National
  Academy of Sciences}\ }\textbf {\bibinfo {volume} {108}},\ \bibinfo {pages}
  {12233} (\bibinfo {year} {2011})}\BibitemShut {NoStop}%
\bibitem [{\citenamefont {Cao}\ \emph {et~al.}(2018{\natexlab{a}})\citenamefont
  {Cao}, \citenamefont {Fatemi}, \citenamefont {Fang}, \citenamefont
  {Watanabe}, \citenamefont {Taniguchi}, \citenamefont {Kaxiras},\ and\
  \citenamefont {Jarillo-Herrero}}]{Cao2018}%
  \BibitemOpen
  \bibfield  {author} {\bibinfo {author} {\bibfnamefont {Y.}~\bibnamefont
  {Cao}}, \bibinfo {author} {\bibfnamefont {V.}~\bibnamefont {Fatemi}},
  \bibinfo {author} {\bibfnamefont {S.}~\bibnamefont {Fang}}, \bibinfo {author}
  {\bibfnamefont {K.}~\bibnamefont {Watanabe}}, \bibinfo {author}
  {\bibfnamefont {T.}~\bibnamefont {Taniguchi}}, \bibinfo {author}
  {\bibfnamefont {E.}~\bibnamefont {Kaxiras}},\ and\ \bibinfo {author}
  {\bibfnamefont {P.}~\bibnamefont {Jarillo-Herrero}},\ }\bibfield  {title}
  {\bibinfo {title} {Unconventional superconductivity in magic-angle graphene
  superlattices},\ }\href {https://doi.org/10.1038/nature26160} {\bibfield
  {journal} {\bibinfo  {journal} {Nature}\ }\textbf {\bibinfo {volume} {556}},\
  \bibinfo {pages} {43} (\bibinfo {year} {2018}{\natexlab{a}})}\BibitemShut
  {NoStop}%
\bibitem [{\citenamefont {Cao}\ \emph {et~al.}(2018{\natexlab{b}})\citenamefont
  {Cao}, \citenamefont {Fatemi}, \citenamefont {Demir}, \citenamefont {Fang},
  \citenamefont {Tomarken}, \citenamefont {Luo}, \citenamefont
  {Sanchez-Yamagishi}, \citenamefont {Watanabe}, \citenamefont {Taniguchi},
  \citenamefont {Kaxiras}, \citenamefont {Ashoori},\ and\ \citenamefont
  {Jarillo-Herrero}}]{Correlated2018}%
  \BibitemOpen
  \bibfield  {author} {\bibinfo {author} {\bibfnamefont {Y.}~\bibnamefont
  {Cao}}, \bibinfo {author} {\bibfnamefont {V.}~\bibnamefont {Fatemi}},
  \bibinfo {author} {\bibfnamefont {A.}~\bibnamefont {Demir}}, \bibinfo
  {author} {\bibfnamefont {S.}~\bibnamefont {Fang}}, \bibinfo {author}
  {\bibfnamefont {S.~L.}\ \bibnamefont {Tomarken}}, \bibinfo {author}
  {\bibfnamefont {J.~Y.}\ \bibnamefont {Luo}}, \bibinfo {author} {\bibfnamefont
  {J.~D.}\ \bibnamefont {Sanchez-Yamagishi}}, \bibinfo {author} {\bibfnamefont
  {K.}~\bibnamefont {Watanabe}}, \bibinfo {author} {\bibfnamefont
  {T.}~\bibnamefont {Taniguchi}}, \bibinfo {author} {\bibfnamefont
  {E.}~\bibnamefont {Kaxiras}}, \bibinfo {author} {\bibfnamefont {R.~C.}\
  \bibnamefont {Ashoori}},\ and\ \bibinfo {author} {\bibfnamefont
  {P.}~\bibnamefont {Jarillo-Herrero}},\ }\bibfield  {title} {\bibinfo {title}
  {Correlated insulator behaviour at half-filling in magic-angle graphene
  superlattices},\ }\href {https://doi.org/10.1038/nature26154} {\bibfield
  {journal} {\bibinfo  {journal} {Nature}\ }\textbf {\bibinfo {volume} {556}},\
  \bibinfo {pages} {80} (\bibinfo {year} {2018}{\natexlab{b}})}\BibitemShut
  {NoStop}%
\bibitem [{\citenamefont {Lu}\ \emph {et~al.}(2019)\citenamefont {Lu},
  \citenamefont {Stepanov}, \citenamefont {Yang}, \citenamefont {Xie},
  \citenamefont {Aamir}, \citenamefont {Das}, \citenamefont {Urgell},
  \citenamefont {Watanabe}, \citenamefont {Taniguchi}, \citenamefont {Zhang},
  \citenamefont {Bachtold}, \citenamefont {MacDonald},\ and\ \citenamefont
  {Efetov}}]{Lu2019}%
  \BibitemOpen
  \bibfield  {author} {\bibinfo {author} {\bibfnamefont {X.}~\bibnamefont
  {Lu}}, \bibinfo {author} {\bibfnamefont {P.}~\bibnamefont {Stepanov}},
  \bibinfo {author} {\bibfnamefont {W.}~\bibnamefont {Yang}}, \bibinfo {author}
  {\bibfnamefont {M.}~\bibnamefont {Xie}}, \bibinfo {author} {\bibfnamefont
  {M.~A.}\ \bibnamefont {Aamir}}, \bibinfo {author} {\bibfnamefont
  {I.}~\bibnamefont {Das}}, \bibinfo {author} {\bibfnamefont {C.}~\bibnamefont
  {Urgell}}, \bibinfo {author} {\bibfnamefont {K.}~\bibnamefont {Watanabe}},
  \bibinfo {author} {\bibfnamefont {T.}~\bibnamefont {Taniguchi}}, \bibinfo
  {author} {\bibfnamefont {G.}~\bibnamefont {Zhang}}, \bibinfo {author}
  {\bibfnamefont {A.}~\bibnamefont {Bachtold}}, \bibinfo {author}
  {\bibfnamefont {A.~H.}\ \bibnamefont {MacDonald}},\ and\ \bibinfo {author}
  {\bibfnamefont {D.~K.}\ \bibnamefont {Efetov}},\ }\bibfield  {title}
  {\bibinfo {title} {Superconductors, orbital magnets and correlated states in
  magic-angle bilayer graphene},\ }\href
  {https://doi.org/10.1038/s41586-019-1695-0} {\bibfield  {journal} {\bibinfo
  {journal} {Nature}\ }\textbf {\bibinfo {volume} {574}},\ \bibinfo {pages}
  {653} (\bibinfo {year} {2019})}\BibitemShut {NoStop}%
\bibitem [{\citenamefont {Sun}\ and\ \citenamefont {Hu}(2020)}]{Sun2020}%
  \BibitemOpen
  \bibfield  {author} {\bibinfo {author} {\bibfnamefont {Z.}~\bibnamefont
  {Sun}}\ and\ \bibinfo {author} {\bibfnamefont {Y.~H.}\ \bibnamefont {Hu}},\
  }\bibfield  {title} {\bibinfo {title} {How magical is magic-angle
  graphene?},\ }\href {https://doi.org/10.1016/j.matt.2020.03.010} {\bibfield
  {journal} {\bibinfo  {journal} {Matter}\ }\textbf {\bibinfo {volume} {2}},\
  \bibinfo {pages} {1106} (\bibinfo {year} {2020})}\BibitemShut {NoStop}%
\bibitem [{\citenamefont {Yankowitz}\ \emph {et~al.}(2019)\citenamefont
  {Yankowitz}, \citenamefont {Chen}, \citenamefont {Polshyn}, \citenamefont
  {Zhang}, \citenamefont {Watanabe}, \citenamefont {Taniguchi}, \citenamefont
  {Graf}, \citenamefont {Young},\ and\ \citenamefont {Dean}}]{Yankowitz2019}%
  \BibitemOpen
  \bibfield  {author} {\bibinfo {author} {\bibfnamefont {M.}~\bibnamefont
  {Yankowitz}}, \bibinfo {author} {\bibfnamefont {S.}~\bibnamefont {Chen}},
  \bibinfo {author} {\bibfnamefont {H.}~\bibnamefont {Polshyn}}, \bibinfo
  {author} {\bibfnamefont {Y.}~\bibnamefont {Zhang}}, \bibinfo {author}
  {\bibfnamefont {K.}~\bibnamefont {Watanabe}}, \bibinfo {author}
  {\bibfnamefont {T.}~\bibnamefont {Taniguchi}}, \bibinfo {author}
  {\bibfnamefont {D.}~\bibnamefont {Graf}}, \bibinfo {author} {\bibfnamefont
  {A.~F.}\ \bibnamefont {Young}},\ and\ \bibinfo {author} {\bibfnamefont
  {C.~R.}\ \bibnamefont {Dean}},\ }\bibfield  {title} {\bibinfo {title} {Tuning
  superconductivity in twisted bilayer graphene},\ }\href
  {https://doi.org/10.1126/science.aav1910} {\bibfield  {journal} {\bibinfo
  {journal} {Science}\ }\textbf {\bibinfo {volume} {363}},\ \bibinfo {pages}
  {1059} (\bibinfo {year} {2019})}\BibitemShut {NoStop}%
\bibitem [{\citenamefont {Carr}\ \emph {et~al.}(2017)\citenamefont {Carr},
  \citenamefont {Massatt}, \citenamefont {Fang}, \citenamefont {Cazeaux},
  \citenamefont {Luskin},\ and\ \citenamefont {Kaxiras}}]{Carr2017}%
  \BibitemOpen
  \bibfield  {author} {\bibinfo {author} {\bibfnamefont {S.}~\bibnamefont
  {Carr}}, \bibinfo {author} {\bibfnamefont {D.}~\bibnamefont {Massatt}},
  \bibinfo {author} {\bibfnamefont {S.}~\bibnamefont {Fang}}, \bibinfo {author}
  {\bibfnamefont {P.}~\bibnamefont {Cazeaux}}, \bibinfo {author} {\bibfnamefont
  {M.}~\bibnamefont {Luskin}},\ and\ \bibinfo {author} {\bibfnamefont
  {E.}~\bibnamefont {Kaxiras}},\ }\bibfield  {title} {\bibinfo {title}
  {Twistronics: Manipulating the electronic properties of two-dimensional
  layered structures through their twist angle},\ }\href
  {https://doi.org/10.1103/PhysRevB.95.075420} {\bibfield  {journal} {\bibinfo
  {journal} {Phys. Rev. B}\ }\textbf {\bibinfo {volume} {95}},\ \bibinfo
  {pages} {075420} (\bibinfo {year} {2017})}\BibitemShut {NoStop}%
\bibitem [{\citenamefont {Tarnopolsky}\ \emph {et~al.}(2019)\citenamefont
  {Tarnopolsky}, \citenamefont {Kruchkov},\ and\ \citenamefont
  {Vishwanath}}]{Tarnopolsky2019}%
  \BibitemOpen
  \bibfield  {author} {\bibinfo {author} {\bibfnamefont {G.}~\bibnamefont
  {Tarnopolsky}}, \bibinfo {author} {\bibfnamefont {A.~J.}\ \bibnamefont
  {Kruchkov}},\ and\ \bibinfo {author} {\bibfnamefont {A.}~\bibnamefont
  {Vishwanath}},\ }\bibfield  {title} {\bibinfo {title} {Origin of magic angles
  in twisted bilayer graphene},\ }\bibfield  {journal} {\bibinfo  {journal}
  {Physical Review Letters}\ }\textbf {\bibinfo {volume} {122}},\ \href
  {https://doi.org/10.1103/physrevlett.122.106405}
  {10.1103/physrevlett.122.106405} (\bibinfo {year} {2019})\BibitemShut
  {NoStop}%
\bibitem [{\citenamefont {Park}\ \emph
  {et~al.}(2008{\natexlab{a}})\citenamefont {Park}, \citenamefont {Yang},
  \citenamefont {Son}, \citenamefont {Cohen},\ and\ \citenamefont
  {Louie}}]{Park2008}%
  \BibitemOpen
  \bibfield  {author} {\bibinfo {author} {\bibfnamefont {C.-H.}\ \bibnamefont
  {Park}}, \bibinfo {author} {\bibfnamefont {L.}~\bibnamefont {Yang}}, \bibinfo
  {author} {\bibfnamefont {Y.-W.}\ \bibnamefont {Son}}, \bibinfo {author}
  {\bibfnamefont {M.~L.}\ \bibnamefont {Cohen}},\ and\ \bibinfo {author}
  {\bibfnamefont {S.~G.}\ \bibnamefont {Louie}},\ }\bibfield  {title} {\bibinfo
  {title} {Anisotropic behaviours of massless dirac~fermions in graphene under
  periodic~potentials},\ }\href {https://doi.org/10.1038/nphys890} {\bibfield
  {journal} {\bibinfo  {journal} {Nature Physics}\ }\textbf {\bibinfo {volume}
  {4}},\ \bibinfo {pages} {213} (\bibinfo {year}
  {2008}{\natexlab{a}})}\BibitemShut {NoStop}%
\bibitem [{\citenamefont {Dubey}\ \emph
  {et~al.}(2013{\natexlab{a}})\citenamefont {Dubey}, \citenamefont {Singh},
  \citenamefont {Bhat}, \citenamefont {Parikh}, \citenamefont {Grover},
  \citenamefont {Sensarma}, \citenamefont {Tripathi}, \citenamefont
  {Sengupta},\ and\ \citenamefont {Deshmukh}}]{Sudipta2013}%
  \BibitemOpen
  \bibfield  {author} {\bibinfo {author} {\bibfnamefont {S.}~\bibnamefont
  {Dubey}}, \bibinfo {author} {\bibfnamefont {V.}~\bibnamefont {Singh}},
  \bibinfo {author} {\bibfnamefont {A.~K.}\ \bibnamefont {Bhat}}, \bibinfo
  {author} {\bibfnamefont {P.}~\bibnamefont {Parikh}}, \bibinfo {author}
  {\bibfnamefont {S.}~\bibnamefont {Grover}}, \bibinfo {author} {\bibfnamefont
  {R.}~\bibnamefont {Sensarma}}, \bibinfo {author} {\bibfnamefont
  {V.}~\bibnamefont {Tripathi}}, \bibinfo {author} {\bibfnamefont
  {K.}~\bibnamefont {Sengupta}},\ and\ \bibinfo {author} {\bibfnamefont
  {M.~M.}\ \bibnamefont {Deshmukh}},\ }\bibfield  {title} {\bibinfo {title}
  {Tunable superlattice in graphene to control the number of dirac points},\
  }\href@noop {} {\bibfield  {journal} {\bibinfo  {journal} {Nano letters}\
  }\textbf {\bibinfo {volume} {13}},\ \bibinfo {pages} {3990} (\bibinfo {year}
  {2013}{\natexlab{a}})}\BibitemShut {NoStop}%
\bibitem [{\citenamefont {Li}\ \emph {et~al.}(2021)\citenamefont {Li},
  \citenamefont {Dietrich}, \citenamefont {Forsythe}, \citenamefont
  {Taniguchi}, \citenamefont {Watanabe}, \citenamefont {Moon},\ and\
  \citenamefont {Dean}}]{Li2021}%
  \BibitemOpen
  \bibfield  {author} {\bibinfo {author} {\bibfnamefont {Y.}~\bibnamefont
  {Li}}, \bibinfo {author} {\bibfnamefont {S.}~\bibnamefont {Dietrich}},
  \bibinfo {author} {\bibfnamefont {C.}~\bibnamefont {Forsythe}}, \bibinfo
  {author} {\bibfnamefont {T.}~\bibnamefont {Taniguchi}}, \bibinfo {author}
  {\bibfnamefont {K.}~\bibnamefont {Watanabe}}, \bibinfo {author}
  {\bibfnamefont {P.}~\bibnamefont {Moon}},\ and\ \bibinfo {author}
  {\bibfnamefont {C.~R.}\ \bibnamefont {Dean}},\ }\bibfield  {title} {\bibinfo
  {title} {Anisotropic band flattening in graphene with one-dimensional
  superlattices},\ }\href {https://doi.org/10.1038/s41565-021-00849-9}
  {\bibfield  {journal} {\bibinfo  {journal} {Nature Nanotechnology}\ }\textbf
  {\bibinfo {volume} {16}},\ \bibinfo {pages} {525} (\bibinfo {year}
  {2021})}\BibitemShut {NoStop}%
\bibitem [{\citenamefont {Park}\ \emph
  {et~al.}(2008{\natexlab{b}})\citenamefont {Park}, \citenamefont {Yang},
  \citenamefont {Son}, \citenamefont {Cohen},\ and\ \citenamefont
  {Louie}}]{CheolHwan2008}%
  \BibitemOpen
  \bibfield  {author} {\bibinfo {author} {\bibfnamefont {C.-H.}\ \bibnamefont
  {Park}}, \bibinfo {author} {\bibfnamefont {L.}~\bibnamefont {Yang}}, \bibinfo
  {author} {\bibfnamefont {Y.-W.}\ \bibnamefont {Son}}, \bibinfo {author}
  {\bibfnamefont {M.~L.}\ \bibnamefont {Cohen}},\ and\ \bibinfo {author}
  {\bibfnamefont {S.~G.}\ \bibnamefont {Louie}},\ }\bibfield  {title} {\bibinfo
  {title} {New generation of massless dirac fermions in graphene under external
  periodic potentials},\ }\href
  {https://doi.org/10.1103/PhysRevLett.101.126804} {\bibfield  {journal}
  {\bibinfo  {journal} {Phys. Rev. Lett.}\ }\textbf {\bibinfo {volume} {101}},\
  \bibinfo {pages} {126804} (\bibinfo {year} {2008}{\natexlab{b}})}\BibitemShut
  {NoStop}%
\bibitem [{\citenamefont {Forsythe}\ \emph {et~al.}(2018)\citenamefont
  {Forsythe}, \citenamefont {Zhou}, \citenamefont {Watanabe}, \citenamefont
  {Taniguchi}, \citenamefont {Pasupathy}, \citenamefont {Moon}, \citenamefont
  {Koshino}, \citenamefont {Kim},\ and\ \citenamefont {Dean}}]{Forsythe2018}%
  \BibitemOpen
  \bibfield  {author} {\bibinfo {author} {\bibfnamefont {C.}~\bibnamefont
  {Forsythe}}, \bibinfo {author} {\bibfnamefont {X.}~\bibnamefont {Zhou}},
  \bibinfo {author} {\bibfnamefont {K.}~\bibnamefont {Watanabe}}, \bibinfo
  {author} {\bibfnamefont {T.}~\bibnamefont {Taniguchi}}, \bibinfo {author}
  {\bibfnamefont {A.}~\bibnamefont {Pasupathy}}, \bibinfo {author}
  {\bibfnamefont {P.}~\bibnamefont {Moon}}, \bibinfo {author} {\bibfnamefont
  {M.}~\bibnamefont {Koshino}}, \bibinfo {author} {\bibfnamefont
  {P.}~\bibnamefont {Kim}},\ and\ \bibinfo {author} {\bibfnamefont {C.~R.}\
  \bibnamefont {Dean}},\ }\bibfield  {title} {\bibinfo {title} {Band structure
  engineering of 2d materials using patterned dielectric superlattices},\
  }\href {https://doi.org/10.1038/s41565-018-0138-7} {\bibfield  {journal}
  {\bibinfo  {journal} {Nature Nanotechnology}\ }\textbf {\bibinfo {volume}
  {13}},\ \bibinfo {pages} {566} (\bibinfo {year} {2018})}\BibitemShut
  {NoStop}%
\bibitem [{\citenamefont {Huang}\ \emph {et~al.}(2015)\citenamefont {Huang},
  \citenamefont {Jnawali}, \citenamefont {Hsu}, \citenamefont {Dhingra},
  \citenamefont {Lee}, \citenamefont {Ryu}, \citenamefont {Bi}, \citenamefont
  {Ghahari}, \citenamefont {Ravichandran}, \citenamefont {Chen} \emph
  {et~al.}}]{huang2015}%
  \BibitemOpen
  \bibfield  {author} {\bibinfo {author} {\bibfnamefont {M.}~\bibnamefont
  {Huang}}, \bibinfo {author} {\bibfnamefont {G.}~\bibnamefont {Jnawali}},
  \bibinfo {author} {\bibfnamefont {J.-F.}\ \bibnamefont {Hsu}}, \bibinfo
  {author} {\bibfnamefont {S.}~\bibnamefont {Dhingra}}, \bibinfo {author}
  {\bibfnamefont {H.}~\bibnamefont {Lee}}, \bibinfo {author} {\bibfnamefont
  {S.}~\bibnamefont {Ryu}}, \bibinfo {author} {\bibfnamefont {F.}~\bibnamefont
  {Bi}}, \bibinfo {author} {\bibfnamefont {F.}~\bibnamefont {Ghahari}},
  \bibinfo {author} {\bibfnamefont {J.}~\bibnamefont {Ravichandran}}, \bibinfo
  {author} {\bibfnamefont {L.}~\bibnamefont {Chen}}, \emph {et~al.},\
  }\bibfield  {title} {\bibinfo {title} {Electric field effects in
  graphene/laalo3/srtio3 heterostructures and nanostructures},\ }\href@noop {}
  {\bibfield  {journal} {\bibinfo  {journal} {APL materials}\ }\textbf
  {\bibinfo {volume} {3}},\ \bibinfo {pages} {062502} (\bibinfo {year}
  {2015})}\BibitemShut {NoStop}%
\bibitem [{\citenamefont {Jnawali}\ \emph {et~al.}(2018)\citenamefont
  {Jnawali}, \citenamefont {Lee}, \citenamefont {Lee}, \citenamefont {Huang},
  \citenamefont {Hsu}, \citenamefont {Bi}, \citenamefont {Zhou}, \citenamefont
  {Cheng}, \citenamefont {D’Urso}, \citenamefont {Irvin} \emph
  {et~al.}}]{jnawali2018}%
  \BibitemOpen
  \bibfield  {author} {\bibinfo {author} {\bibfnamefont {G.}~\bibnamefont
  {Jnawali}}, \bibinfo {author} {\bibfnamefont {H.}~\bibnamefont {Lee}},
  \bibinfo {author} {\bibfnamefont {J.-W.}\ \bibnamefont {Lee}}, \bibinfo
  {author} {\bibfnamefont {M.}~\bibnamefont {Huang}}, \bibinfo {author}
  {\bibfnamefont {J.-F.}\ \bibnamefont {Hsu}}, \bibinfo {author} {\bibfnamefont
  {F.}~\bibnamefont {Bi}}, \bibinfo {author} {\bibfnamefont {R.}~\bibnamefont
  {Zhou}}, \bibinfo {author} {\bibfnamefont {G.}~\bibnamefont {Cheng}},
  \bibinfo {author} {\bibfnamefont {B.}~\bibnamefont {D’Urso}}, \bibinfo
  {author} {\bibfnamefont {P.}~\bibnamefont {Irvin}}, \emph {et~al.},\
  }\bibfield  {title} {\bibinfo {title} {Graphene-complex-oxide nanoscale
  device concepts},\ }\href@noop {} {\bibfield  {journal} {\bibinfo  {journal}
  {ACS nano}\ }\textbf {\bibinfo {volume} {12}},\ \bibinfo {pages} {6128}
  (\bibinfo {year} {2018})}\BibitemShut {NoStop}%
\bibitem [{\citenamefont {Yang}\ \emph {et~al.}(2020)\citenamefont {Yang},
  \citenamefont {Hao}, \citenamefont {Chen}, \citenamefont {Guo}, \citenamefont
  {Yu}, \citenamefont {Hu}, \citenamefont {Eom}, \citenamefont {Lee},
  \citenamefont {Eom}, \citenamefont {Irvin} \emph {et~al.}}]{yang2020}%
  \BibitemOpen
  \bibfield  {author} {\bibinfo {author} {\bibfnamefont {D.}~\bibnamefont
  {Yang}}, \bibinfo {author} {\bibfnamefont {S.}~\bibnamefont {Hao}}, \bibinfo
  {author} {\bibfnamefont {J.}~\bibnamefont {Chen}}, \bibinfo {author}
  {\bibfnamefont {Q.}~\bibnamefont {Guo}}, \bibinfo {author} {\bibfnamefont
  {M.}~\bibnamefont {Yu}}, \bibinfo {author} {\bibfnamefont {Y.}~\bibnamefont
  {Hu}}, \bibinfo {author} {\bibfnamefont {K.}~\bibnamefont {Eom}}, \bibinfo
  {author} {\bibfnamefont {J.-W.}\ \bibnamefont {Lee}}, \bibinfo {author}
  {\bibfnamefont {C.-B.}\ \bibnamefont {Eom}}, \bibinfo {author} {\bibfnamefont
  {P.}~\bibnamefont {Irvin}}, \emph {et~al.},\ }\bibfield  {title} {\bibinfo
  {title} {Nanoscale control of laalo3/srtio3 metal--insulator transition using
  ultra-low-voltage electron-beam lithography},\ }\href@noop {} {\bibfield
  {journal} {\bibinfo  {journal} {Applied Physics Letters}\ }\textbf {\bibinfo
  {volume} {117}},\ \bibinfo {pages} {253103} (\bibinfo {year}
  {2020})}\BibitemShut {NoStop}%
\bibitem [{\citenamefont {Zou}\ \emph {et~al.}(2018)\citenamefont {Zou},
  \citenamefont {Po}, \citenamefont {Vishwanath},\ and\ \citenamefont
  {Senthil}}]{Zou2018}%
  \BibitemOpen
  \bibfield  {author} {\bibinfo {author} {\bibfnamefont {L.}~\bibnamefont
  {Zou}}, \bibinfo {author} {\bibfnamefont {H.~C.}\ \bibnamefont {Po}},
  \bibinfo {author} {\bibfnamefont {A.}~\bibnamefont {Vishwanath}},\ and\
  \bibinfo {author} {\bibfnamefont {T.}~\bibnamefont {Senthil}},\ }\bibfield
  {title} {\bibinfo {title} {Band structure of twisted bilayer graphene:
  Emergent symmetries, commensurate approximants, and wannier obstructions},\
  }\bibfield  {journal} {\bibinfo  {journal} {Physical Review B}\ }\textbf
  {\bibinfo {volume} {98}},\ \href {https://doi.org/10.1103/physrevb.98.085435}
  {10.1103/physrevb.98.085435} (\bibinfo {year} {2018})\BibitemShut {NoStop}%
\bibitem [{\citenamefont {McCann}\ and\ \citenamefont
  {Koshino}(2013)}]{McCann_2013}%
  \BibitemOpen
  \bibfield  {author} {\bibinfo {author} {\bibfnamefont {E.}~\bibnamefont
  {McCann}}\ and\ \bibinfo {author} {\bibfnamefont {M.}~\bibnamefont
  {Koshino}},\ }\bibfield  {title} {\bibinfo {title} {The electronic properties
  of bilayer graphene},\ }\href {https://doi.org/10.1088/0034-4885/76/5/056503}
  {\bibfield  {journal} {\bibinfo  {journal} {Reports on Progress in Physics}\
  }\textbf {\bibinfo {volume} {76}},\ \bibinfo {pages} {056503} (\bibinfo
  {year} {2013})}\BibitemShut {NoStop}%
\bibitem [{\citenamefont {Neto}\ \emph {et~al.}(2009)\citenamefont {Neto},
  \citenamefont {Guinea}, \citenamefont {Peres}, \citenamefont {Novoselov},\
  and\ \citenamefont {Geim}}]{CastroNeto2009}%
  \BibitemOpen
  \bibfield  {author} {\bibinfo {author} {\bibfnamefont {A.~H.~C.}\
  \bibnamefont {Neto}}, \bibinfo {author} {\bibfnamefont {F.}~\bibnamefont
  {Guinea}}, \bibinfo {author} {\bibfnamefont {N.~M.~R.}\ \bibnamefont
  {Peres}}, \bibinfo {author} {\bibfnamefont {K.~S.}\ \bibnamefont
  {Novoselov}},\ and\ \bibinfo {author} {\bibfnamefont {A.~K.}\ \bibnamefont
  {Geim}},\ }\bibfield  {title} {\bibinfo {title} {The electronic properties of
  graphene},\ }\href {https://doi.org/10.1103/revmodphys.81.109} {\bibfield
  {journal} {\bibinfo  {journal} {Reviews of Modern Physics}\ }\textbf
  {\bibinfo {volume} {81}},\ \bibinfo {pages} {109} (\bibinfo {year}
  {2009})}\BibitemShut {NoStop}%
\bibitem [{\citenamefont {Andrei}\ and\ \citenamefont
  {MacDonald}(2020)}]{Andrei2020}%
  \BibitemOpen
  \bibfield  {author} {\bibinfo {author} {\bibfnamefont {E.~Y.}\ \bibnamefont
  {Andrei}}\ and\ \bibinfo {author} {\bibfnamefont {A.~H.}\ \bibnamefont
  {MacDonald}},\ }\bibfield  {title} {\bibinfo {title} {Graphene bilayers with
  a twist},\ }\href@noop {} {\bibfield  {journal} {\bibinfo  {journal} {Nature
  materials}\ }\textbf {\bibinfo {volume} {19}},\ \bibinfo {pages} {1265}
  (\bibinfo {year} {2020})}\BibitemShut {NoStop}%
\bibitem [{\citenamefont {Po}\ \emph {et~al.}(2018)\citenamefont {Po},
  \citenamefont {Zou}, \citenamefont {Vishwanath},\ and\ \citenamefont
  {Senthil}}]{Po2018}%
  \BibitemOpen
  \bibfield  {author} {\bibinfo {author} {\bibfnamefont {H.~C.}\ \bibnamefont
  {Po}}, \bibinfo {author} {\bibfnamefont {L.}~\bibnamefont {Zou}}, \bibinfo
  {author} {\bibfnamefont {A.}~\bibnamefont {Vishwanath}},\ and\ \bibinfo
  {author} {\bibfnamefont {T.}~\bibnamefont {Senthil}},\ }\bibfield  {title}
  {\bibinfo {title} {Origin of mott insulating behavior and superconductivity
  in twisted bilayer graphene},\ }\bibfield  {journal} {\bibinfo  {journal}
  {Physical Review X}\ }\textbf {\bibinfo {volume} {8}},\ \href
  {https://doi.org/10.1103/physrevx.8.031089} {10.1103/physrevx.8.031089}
  (\bibinfo {year} {2018})\BibitemShut {NoStop}%
\bibitem [{\citenamefont {Po}\ \emph {et~al.}(2019)\citenamefont {Po},
  \citenamefont {Zou}, \citenamefont {Senthil},\ and\ \citenamefont
  {Vishwanath}}]{Po2019}%
  \BibitemOpen
  \bibfield  {author} {\bibinfo {author} {\bibfnamefont {H.~C.}\ \bibnamefont
  {Po}}, \bibinfo {author} {\bibfnamefont {L.}~\bibnamefont {Zou}}, \bibinfo
  {author} {\bibfnamefont {T.}~\bibnamefont {Senthil}},\ and\ \bibinfo {author}
  {\bibfnamefont {A.}~\bibnamefont {Vishwanath}},\ }\bibfield  {title}
  {\bibinfo {title} {Faithful tight-binding models and fragile topology of
  magic-angle bilayer graphene},\ }\href
  {https://doi.org/10.1103/PhysRevB.99.195455} {\bibfield  {journal} {\bibinfo
  {journal} {Phys. Rev. B}\ }\textbf {\bibinfo {volume} {99}},\ \bibinfo
  {pages} {195455} (\bibinfo {year} {2019})}\BibitemShut {NoStop}%
\bibitem [{\citenamefont {Shi}\ \emph {et~al.}(2019)\citenamefont {Shi},
  \citenamefont {Ma},\ and\ \citenamefont {Song}}]{Shi2019}%
  \BibitemOpen
  \bibfield  {author} {\bibinfo {author} {\bibfnamefont {L.}~\bibnamefont
  {Shi}}, \bibinfo {author} {\bibfnamefont {J.}~\bibnamefont {Ma}},\ and\
  \bibinfo {author} {\bibfnamefont {J.~C.~W.}\ \bibnamefont {Song}},\
  }\bibfield  {title} {\bibinfo {title} {Gate-tunable flat bands in van der
  waals patterned dielectric superlattices},\ }\href
  {https://doi.org/10.1088/2053-1583/ab59a8} {\bibfield  {journal} {\bibinfo
  {journal} {2D Materials}\ }\textbf {\bibinfo {volume} {7}},\ \bibinfo {pages}
  {015028} (\bibinfo {year} {2019})}\BibitemShut {NoStop}%
\bibitem [{\citenamefont {Chaves}\ \emph {et~al.}(2020)\citenamefont {Chaves},
  \citenamefont {Azadani}, \citenamefont {Alsalman}, \citenamefont {da~Costa},
  \citenamefont {Frisenda}, \citenamefont {Chaves}, \citenamefont {Song},
  \citenamefont {Kim}, \citenamefont {He}, \citenamefont {Zhou}, \citenamefont
  {Castellanos-Gomez}, \citenamefont {Peeters}, \citenamefont {Liu},
  \citenamefont {Hinkle}, \citenamefont {Oh}, \citenamefont {Ye}, \citenamefont
  {Koester}, \citenamefont {Lee}, \citenamefont {Avouris}, \citenamefont
  {Wang},\ and\ \citenamefont {Low}}]{Chaves2020}%
  \BibitemOpen
  \bibfield  {author} {\bibinfo {author} {\bibfnamefont {A.}~\bibnamefont
  {Chaves}}, \bibinfo {author} {\bibfnamefont {J.~G.}\ \bibnamefont {Azadani}},
  \bibinfo {author} {\bibfnamefont {H.}~\bibnamefont {Alsalman}}, \bibinfo
  {author} {\bibfnamefont {D.~R.}\ \bibnamefont {da~Costa}}, \bibinfo {author}
  {\bibfnamefont {R.}~\bibnamefont {Frisenda}}, \bibinfo {author}
  {\bibfnamefont {A.~J.}\ \bibnamefont {Chaves}}, \bibinfo {author}
  {\bibfnamefont {S.~H.}\ \bibnamefont {Song}}, \bibinfo {author}
  {\bibfnamefont {Y.~D.}\ \bibnamefont {Kim}}, \bibinfo {author} {\bibfnamefont
  {D.}~\bibnamefont {He}}, \bibinfo {author} {\bibfnamefont {J.}~\bibnamefont
  {Zhou}}, \bibinfo {author} {\bibfnamefont {A.}~\bibnamefont
  {Castellanos-Gomez}}, \bibinfo {author} {\bibfnamefont {F.~M.}\ \bibnamefont
  {Peeters}}, \bibinfo {author} {\bibfnamefont {Z.}~\bibnamefont {Liu}},
  \bibinfo {author} {\bibfnamefont {C.~L.}\ \bibnamefont {Hinkle}}, \bibinfo
  {author} {\bibfnamefont {S.-H.}\ \bibnamefont {Oh}}, \bibinfo {author}
  {\bibfnamefont {P.~D.}\ \bibnamefont {Ye}}, \bibinfo {author} {\bibfnamefont
  {S.~J.}\ \bibnamefont {Koester}}, \bibinfo {author} {\bibfnamefont {Y.~H.}\
  \bibnamefont {Lee}}, \bibinfo {author} {\bibfnamefont {P.}~\bibnamefont
  {Avouris}}, \bibinfo {author} {\bibfnamefont {X.}~\bibnamefont {Wang}},\ and\
  \bibinfo {author} {\bibfnamefont {T.}~\bibnamefont {Low}},\ }\bibfield
  {title} {\bibinfo {title} {Bandgap engineering of two-dimensional
  semiconductor materials},\ }\bibfield  {journal} {\bibinfo  {journal} {npj 2d
  materials and applications}\ }\textbf {\bibinfo {volume} {4}},\ \href
  {https://doi.org/10.1038/s41699-020-00162-4} {10.1038/s41699-020-00162-4}
  (\bibinfo {year} {2020})\BibitemShut {NoStop}%
\bibitem [{\citenamefont {Bergman}\ \emph {et~al.}(2008)\citenamefont
  {Bergman}, \citenamefont {Wu},\ and\ \citenamefont {Balents}}]{Bergman2008}%
  \BibitemOpen
  \bibfield  {author} {\bibinfo {author} {\bibfnamefont {D.~L.}\ \bibnamefont
  {Bergman}}, \bibinfo {author} {\bibfnamefont {C.}~\bibnamefont {Wu}},\ and\
  \bibinfo {author} {\bibfnamefont {L.}~\bibnamefont {Balents}},\ }\bibfield
  {title} {\bibinfo {title} {Band touching from real-space topology in
  frustrated hopping models},\ }\bibfield  {journal} {\bibinfo  {journal}
  {Physical Review B}\ }\textbf {\bibinfo {volume} {78}},\ \href
  {https://doi.org/10.1103/physrevb.78.125104} {10.1103/physrevb.78.125104}
  (\bibinfo {year} {2008})\BibitemShut {NoStop}%
\bibitem [{\citenamefont {Cen}\ \emph {et~al.}(2008)\citenamefont {Cen},
  \citenamefont {Thiel}, \citenamefont {Hammerl}, \citenamefont {Schneider},
  \citenamefont {Andersen}, \citenamefont {Hellberg}, \citenamefont
  {Mannhart},\ and\ \citenamefont {Levy}}]{cen2008}%
  \BibitemOpen
  \bibfield  {author} {\bibinfo {author} {\bibfnamefont {C.}~\bibnamefont
  {Cen}}, \bibinfo {author} {\bibfnamefont {S.}~\bibnamefont {Thiel}}, \bibinfo
  {author} {\bibfnamefont {G.}~\bibnamefont {Hammerl}}, \bibinfo {author}
  {\bibfnamefont {C.~W.}\ \bibnamefont {Schneider}}, \bibinfo {author}
  {\bibfnamefont {K.}~\bibnamefont {Andersen}}, \bibinfo {author}
  {\bibfnamefont {C.~S.}\ \bibnamefont {Hellberg}}, \bibinfo {author}
  {\bibfnamefont {J.}~\bibnamefont {Mannhart}},\ and\ \bibinfo {author}
  {\bibfnamefont {J.}~\bibnamefont {Levy}},\ }\bibfield  {title} {\bibinfo
  {title} {Nanoscale control of an interfacial metal--insulator transition at
  room temperature},\ }\href@noop {} {\bibfield  {journal} {\bibinfo  {journal}
  {Nature materials}\ }\textbf {\bibinfo {volume} {7}},\ \bibinfo {pages} {298}
  (\bibinfo {year} {2008})}\BibitemShut {NoStop}%
\bibitem [{\citenamefont {Park}\ \emph
  {et~al.}(2008{\natexlab{c}})\citenamefont {Park}, \citenamefont {Yang},
  \citenamefont {Son}, \citenamefont {Cohen},\ and\ \citenamefont
  {Louie}}]{Park2008a}%
  \BibitemOpen
  \bibfield  {author} {\bibinfo {author} {\bibfnamefont {C.-H.}\ \bibnamefont
  {Park}}, \bibinfo {author} {\bibfnamefont {L.}~\bibnamefont {Yang}}, \bibinfo
  {author} {\bibfnamefont {Y.-W.}\ \bibnamefont {Son}}, \bibinfo {author}
  {\bibfnamefont {M.~L.}\ \bibnamefont {Cohen}},\ and\ \bibinfo {author}
  {\bibfnamefont {S.~G.}\ \bibnamefont {Louie}},\ }\bibfield  {title} {\bibinfo
  {title} {New generation of massless dirac fermions in graphene under external
  periodic potentials},\ }\bibfield  {journal} {\bibinfo  {journal} {Physical
  Review Letters}\ }\textbf {\bibinfo {volume} {101}},\ \href
  {https://doi.org/10.1103/physrevlett.101.126804}
  {10.1103/physrevlett.101.126804} (\bibinfo {year}
  {2008}{\natexlab{c}})\BibitemShut {NoStop}%
\bibitem [{\citenamefont {Park}\ \emph {et~al.}(2009)\citenamefont {Park},
  \citenamefont {Son}, \citenamefont {Yang}, \citenamefont {Cohen},\ and\
  \citenamefont {Louie}}]{Park2009}%
  \BibitemOpen
  \bibfield  {author} {\bibinfo {author} {\bibfnamefont {C.-H.}\ \bibnamefont
  {Park}}, \bibinfo {author} {\bibfnamefont {Y.-W.}\ \bibnamefont {Son}},
  \bibinfo {author} {\bibfnamefont {L.}~\bibnamefont {Yang}}, \bibinfo {author}
  {\bibfnamefont {M.~L.}\ \bibnamefont {Cohen}},\ and\ \bibinfo {author}
  {\bibfnamefont {S.~G.}\ \bibnamefont {Louie}},\ }\bibfield  {title} {\bibinfo
  {title} {Landau levels and quantum hall effect in graphene superlattices},\
  }\bibfield  {journal} {\bibinfo  {journal} {Physical Review Letters}\
  }\textbf {\bibinfo {volume} {103}},\ \href
  {https://doi.org/10.1103/physrevlett.103.046808}
  {10.1103/physrevlett.103.046808} (\bibinfo {year} {2009})\BibitemShut
  {NoStop}%
\bibitem [{\citenamefont {Brey}\ and\ \citenamefont {Fertig}(2009)}]{Brey2009}%
  \BibitemOpen
  \bibfield  {author} {\bibinfo {author} {\bibfnamefont {L.}~\bibnamefont
  {Brey}}\ and\ \bibinfo {author} {\bibfnamefont {H.~A.}\ \bibnamefont
  {Fertig}},\ }\bibfield  {title} {\bibinfo {title} {Emerging zero modes for
  graphene in a periodic potential},\ }\bibfield  {journal} {\bibinfo
  {journal} {Physical Review Letters}\ }\textbf {\bibinfo {volume} {103}},\
  \href {https://doi.org/10.1103/physrevlett.103.046809}
  {10.1103/physrevlett.103.046809} (\bibinfo {year} {2009})\BibitemShut
  {NoStop}%
\bibitem [{\citenamefont {Barbier}\ \emph {et~al.}(2010)\citenamefont
  {Barbier}, \citenamefont {Vasilopoulos},\ and\ \citenamefont
  {Peeters}}]{Barbier2010}%
  \BibitemOpen
  \bibfield  {author} {\bibinfo {author} {\bibfnamefont {M.}~\bibnamefont
  {Barbier}}, \bibinfo {author} {\bibfnamefont {P.}~\bibnamefont
  {Vasilopoulos}},\ and\ \bibinfo {author} {\bibfnamefont {F.~M.}\ \bibnamefont
  {Peeters}},\ }\bibfield  {title} {\bibinfo {title} {Extra dirac points in the
  energy spectrum for superlattices on single-layer graphene},\ }\bibfield
  {journal} {\bibinfo  {journal} {Physical Review B}\ }\textbf {\bibinfo
  {volume} {81}},\ \href {https://doi.org/10.1103/physrevb.81.075438}
  {10.1103/physrevb.81.075438} (\bibinfo {year} {2010})\BibitemShut {NoStop}%
\bibitem [{\citenamefont {Dubey}\ \emph
  {et~al.}(2013{\natexlab{b}})\citenamefont {Dubey}, \citenamefont {Singh},
  \citenamefont {Bhat}, \citenamefont {Parikh}, \citenamefont {Grover},
  \citenamefont {Sensarma}, \citenamefont {Tripathi}, \citenamefont
  {Sengupta},\ and\ \citenamefont {Deshmukh}}]{Dubey2013}%
  \BibitemOpen
  \bibfield  {author} {\bibinfo {author} {\bibfnamefont {S.}~\bibnamefont
  {Dubey}}, \bibinfo {author} {\bibfnamefont {V.}~\bibnamefont {Singh}},
  \bibinfo {author} {\bibfnamefont {A.~K.}\ \bibnamefont {Bhat}}, \bibinfo
  {author} {\bibfnamefont {P.}~\bibnamefont {Parikh}}, \bibinfo {author}
  {\bibfnamefont {S.}~\bibnamefont {Grover}}, \bibinfo {author} {\bibfnamefont
  {R.}~\bibnamefont {Sensarma}}, \bibinfo {author} {\bibfnamefont
  {V.}~\bibnamefont {Tripathi}}, \bibinfo {author} {\bibfnamefont
  {K.}~\bibnamefont {Sengupta}},\ and\ \bibinfo {author} {\bibfnamefont
  {M.~M.}\ \bibnamefont {Deshmukh}},\ }\bibfield  {title} {\bibinfo {title}
  {Tunable superlattice in graphene to control the number of dirac points},\
  }\href {https://doi.org/10.1021/nl4006029} {\bibfield  {journal} {\bibinfo
  {journal} {Nano Letters}\ }\textbf {\bibinfo {volume} {13}},\ \bibinfo
  {pages} {3990} (\bibinfo {year} {2013}{\natexlab{b}})}\BibitemShut {NoStop}%
\end{thebibliography}%


\end{document}